\documentclass[12pt]{article}
\pdfoutput=1
\usepackage{latexsym}
\usepackage{amsmath,,calrsfs}
\usepackage{amsfonts}
\usepackage{amssymb}
\usepackage{amscd}
\usepackage{bbm}
\usepackage{fancybox}
\usepackage{cite}
\usepackage{amsmath,amsfonts,amsbsy}
\usepackage{pstricks,pst-node}
\usepackage[small,bf,hang]{caption2}
\usepackage{graphicx}
\usepackage{epsfig}
\usepackage{psfrag}
\usepackage{comment}

\usepackage{float}

\psset{unit=1.3cm,linewidth=.5pt,radius=.2}  

\usepackage{multirow}                     
\usepackage{float}                          
\usepackage{lscape}                         
\usepackage{bm}

\newcommand{\bi}{\begin{itemize}}
\newcommand{\ei}{\end{itemize}}
\newcommand{\bea}{\begin{eqnarray}}
\newcommand{\eea}{\end{eqnarray}}
\newcommand{\bt}{\begin{tabular}}
\newcommand{\et}{\end{tabular}}
\newcommand{\bc}{\begin{center}}
\newcommand{\ec}{\end{center}}

\def\theequation{\arabic{section}.\arabic{equation}}

\newcommand{\be}{\begin{equation}}
\newcommand{\ee}{\end{equation}}
\newcommand{\ba}{\begin{array}}
\newcommand{\ea}{\end{array}}
\newcommand{\p}[1]{(\ref{#1})}
\newcommand{\lb}[1]{\label{#1}}

\def\bbox{{\,\lower0.9pt\vbox{\hrule \hbox{\vrule height 0.2 cm
\hskip 0.2 cm \vrule height 0.2 cm}\hrule}\,}}
\newcommand{\dsl}{\pa \kern-0.5em /}

\newcommand{\nn}{\nonumber \\}

\makeatletter \@addtoreset{equation}{section} \makeatother
\renewcommand{\theequation}{\thesection.\arabic{equation}}

\def\slashchar#1{\setbox0=\hbox{$#1$}           
   \dimen0=\wd0                                 
   \setbox1=\hbox{/} \dimen1=\wd1               
   \ifdim\dimen0>\dimen1                        
      \rlap{\hbox to \dimen0{\hfil/\hfil}}      
      #1                                        
   \else                                        
      \rlap{\hbox to \dimen1{\hfil$#1$\hfil}}   
      /                                         
   \fi}

\begin{document}

\begin{titlepage}

\renewcommand{\thefootnote}{\star}

\begin{center}

\hfill  {}


{\Large \bf  Higher-dimensional invariants in
\vspace{0.2cm}

$6D$ super Yang-Mills theory}
\vspace{0.3cm}

\vspace{1.3cm}
\renewcommand{\thefootnote}{$\star$}

{\large\bf Serafim Buyucli} ${}^\dag$,
{\quad \large\bf Evgeny~Ivanov} ${}^{\ast,\, \star}$
 \vspace{0.5cm}

{${}^\ast$ \it Bogoliubov Laboratory of Theoretical Physics, JINR,}\\
{\it 141980 Dubna, Moscow region, Russia} \\

{${}^\star$ \it Moscow Institute of Physics and Technology,}\\
{\it 141700 Dolgoprudny, Moscow region, Russia}
\vspace{0.1cm}

{\tt eivanov@theor.jinr.ru}\\
\vspace{0.4cm}

\end{center}
\vspace{0.2cm} \vskip 0.6truecm \nopagebreak

\begin{abstract}
\noindent We exploit the $6D, {\cal N}=(1,0)$ and ${\cal N}=(1,1)$  harmonic superspace approaches
to construct the full set of the maximally supersymmetric on-shell invariants of the canonical dimension ${\bf d=12}$ in $6D, \,{\cal N}=(1,1)$ supersymmetric Yang-Mills
(SYM) theory. Both single- and double-trace invariants are derived. Only four single-trace
and two double-trace invariants prove to be independent. The invariants constructed can provide the possible counterterms of ${\cal N}=(1,1)$ SYM theory
at four-loop order, where the first
double-trace divergences are expected to appear. We explicitly exhibit the gauge sector of all invariants in terms of ${\cal N} = (1, 0)$
gauge superfields and find the absence of ${\cal N}=(1,1)$ supercompletion of the $F^6$ term in the abelian limit.

\end{abstract}
\vskip 1cm

\vskip 0.5cm

\vspace{1cm}

\phantom{Keywords: }
\vspace{1cm}

{${}^\dag$\it Deceased}\\

\newpage

\end{titlepage}

\setcounter{footnote}{0}

\newpage
\setcounter{page}{1}

\section{Introduction}
Maximally supersymmetric theories in diverse dimensions are of particular interest in view of their close relation to superstrings
and D-branes. This concerns, e.g.,  supersymmetric Yang-Mills (SYM) theories with sixteen supercharges existing in dimensions $D\leq 10$.
In particular, maximally supersymmetric $6D, {\cal N}=(1,1)$  SYM theory arises
as a low-energy effective field theory on coincident D5-branes and is reduced to $4D, {\cal N} = 4$
SYM theory upon the appropriate compactification $6D \rightarrow 4D$. The study of quantum $6D, {\cal N}=(1,1)$  SYM theory attracts a heightened
attention despite the fact that this theory is non-renormalizable by power-counting because of dimensionful coupling constant. The relevant full quantum
effective action is expected to provide an ultra violet (UV) completion of this theory and to be identical, after summing up all loop contributions,
to the full Born-Infeld-type world-volume action of D5-branes \cite{Tseytlin}.

The quantum structure of the maximally extended gauge and supergravity theories is uncovered by studying their amplitudes, for which
various powerful methods have been developed, see, e.g., refs. \cite{3,4,5,6,7,9,13,14,Bern:2012uf,BGRV,Bjornsson:2010wm,Bjornsson:2010wu,BKKTV,25,Bossard:2009mn}
(a more  complete list of references can be found in the paper \cite{25}).
Using these methods, some surprising  features of $6D, {\cal N}=(1,1)$  SYM theory in the quantum domain were observed.
For instance, it is free of logarithmic divergences up to two loops, at three loops such divergences are seen only in the sector
of planar (single-trace) diagrams,
while the non-planar (double-trace) diagrams are finite.

No off-shell superfield formalism for $6D, {\cal N}=(1, 1)$ SYM theory is known, while such a formulation was developed for  $6D, {\cal N}=(1, 0)$
SYM theory in \cite{16,Koller} in the framework of the conventional $6D, {\cal N}=(1, 0)$ superspace. The harmonic superspace (HSS) approach suggested earlier for
$4D, {\cal N}=2$ theories \cite{20,18} was generalized to six dimensions in ref. \cite{19,22}. The higher-dimensional counterterms of ${\cal N}=(1, 0)$ and ${\cal N}=(1, 1)$
SYM theories were studied in conventional $6D, {\cal N}=(1, 0)$ superspace \cite{11,11a} and in $6D$ HSS, as well as in its on-shell extension,
bi-HSS with two sets of $SU(2)$ harmonic variables \cite{Int,25,12} \footnote{A similar bi-harmonic on-shell formulation of $4D, {\cal N}=4$ SYM was recently worked out in \cite{BuIvIv}.}.

The quantum (Wilsonian) effective action of $6D, {\cal N}=(1, 1)$ SYM theory probably involves an infinite set of
counterterms with the higher-order derivatives. Generically, they are not
supersymmetric off shell and need not to be supersymmetric even on the shell of original equations of motion.
Since every new term added to the original action can be looked upon as a deformation of the latter,
the form of the ``on-shell'' supersymmetry transformations can vary from one loop order to another. It is the effective action with the whole set of counterterms
that is expected to possess the properly realized deformed ${\cal N}=(1, 1)$ supersymmetry. As was argued in \cite{Smi}, up to the fifth order in loops
it is legitimate to deal with the transformations in which the  superfields involved are on the shell of the equations of motion corresponding to the original
``microscopic'' action. The deformation effects would manifest themselves only in the next orders. The corresponding canonical (scaling) dimensions
of the component Lagrangians are even and can vary from ${\bf d=6}$ (one loop) to ${\bf d=14}$ (five loops).

Possible higher-dimensional counterterms were explicitly constructed in \cite{12} using the off-shell $6D, {\cal N}=(1, 0)$ and
the on-shell ${\cal N} = (1, 1)$ harmonic superspace approaches. For the canonical dimension ${\bf d = 6}$, all $6D, {\cal N}=(1, 1)$
invariants were shown to vanish on shell, which amounts to the one-loop finiteness  of ${\cal N}=(1, 1)$ SYM theory. The non-trivial
on-shell ${\cal N}=(1, 1)$ supersymmetric and gauge invariant  ${\bf d=8}$ operators exist, though they do not possess the full off-shell
${\cal N} = (1, 0)$ supersymmetry required by the perturbation theory in $6D, {\cal N}=(1,0)$ HSS \footnote{The manifestly off-shell
$6D, {\cal N}=(1,0)$ supersymmetric and gauge covariant quantum superfield technique in $6D$ HSS was worked out and applied in \cite{Q6D1, Q6D2, Q6D3}.}.
Hence, the theory is UV finite up to two loops and UV divergences should start from three loops, in accord with the general reasoning
of \cite{BKKTV}.  The invariants, corresponding to the three-loop order of the perturbation
expansion and having dimension ${\bf d=10}$, are of the two types, planar (single-trace) and non-planar
(double-trace). An interesting property of the single-trace invariant is that it can be equivalently written as an integral over the
full bi-HSS, as well as over  its $3/4$  analytic subspace. The double-trace invariant can be defined only in the analytic subspace.
This may be one of the reasons why the double-trace invariant is UV protected and does not make
contribution to the three-loop amplitude that was revealed by
explicit calculations in \cite{13,14,Bern:2012uf,BGRV,Bjornsson:2010wm,Bjornsson:2010wu}.

In this paper we continue the analysis of ref. \cite{12} and construct the
maximally supersymmetric on-shell invariants of $6D, {\cal N}=(1, 1)$ SYM theory with the canonical dimension ${\bf d = 12}$ which should be responsible
for the four-loop divergences. We find that there are four single-trace invariants and only two double-trace
invariants of this dimension. Unlike the ${\bf d = 10}$
invariants, it is impossible to equivalently rewrite any of the ${\bf d = 12}$ ones as
integrals over the analytic subspaces of $6D, {\cal N}=(1, 1)$ HSS, with the proper manifestly analytic Lagrangian densities.

In Sect. 2 we start with recalling basic features of the spinor $6D$ formalism and of ${\cal N}=(1,0)$ HSS with
one set of $SU(2)$ harmonic variables  and ${\cal N}=(1,1)$ HSS with two sets of  such variables. The brief exposition of the HSS description
of ${\cal N}=(1,0)$ SYM theory and $6D$ hypermultiplets is a subject of Sect. 3. The on-shell constraints of ${\cal N}=(1,1)$ SYM theory in the bi-HSS
and their solution in terms of ${\cal N}=(1,0)$ harmonic superfields found in \cite{12} are reviewed in Sect. 4. In Sect. 5 we recall the
${\cal N}=(1,1)$ HSS form of the on-shell invariants constructed in \cite{12} up to the dimension ${\bf d=10}$. We present a new double-trace ${\bf d=10}$
invariant in which both ${\cal N}=(1,0)$ and ${\cal N}=(0,1)$ supersymmetries are realized on shell. The structure of new ${\bf d=12}$ ${\cal N}=(1,1)$ invariants is
discussed in detail in Sect. 6. In Sect. 7 we consider the reduction of these invariants to ${\cal N}=(1,0)$ SYM invariants. We also study the abelian limit of the
new invariants and show that exists no abelian ${\cal N}=(1,1)$ completion  of the bosonic $F^6$ invariant. Some technical details are brought in Appendix.

This paper is partially based on the bachelor thesis of  Serafim Buyucli, the fourth-grade student of MIPT,  defended by him in JINR under the supervision of E. Ivanov
in June of 2018. Serafim had died tragically in August of 2018.

\section{$6D$ spinor algebra and HSS}

We start by recalling generalities of $6D$ spinor algebra, as well as of ${\cal N} = (1, 0)$  and ${\cal N} = (1, 1)$ ordinary
and harmonic superspaces (HSS).
More details can be found in \cite{16,12,17,20,18,19,22}. We will basically follow the notation and conventions of \cite{12}.

The group $Spin(5,1)$  has two inequivalent irreducible spinor representations of complex
dimension  4, that is $(1, 0)$ and $(0, 1)$ spinors. In contrast to $4D$ case, the complex
conjugation does not interchange them, but produces an equivalent representation.
In what follows, we will deal with the symplectic Majorana-Weyl spinors
$\lambda^a_i$ and $\psi_{aA}$ carrying extra $SU(2) \simeq Sp(1)$
indices $i =1,2; A=1,2$ and satisfying the {\it pseudoreality}
conditions, e.g.,
\be
\overline{\lambda^a_i} = -C^a_{\,\,b}
(\lambda^b_i)^* = \varepsilon^{ik}\lambda^a_k\,,
\ee
$C$ being the charge conjugation matrix ($C =-C^T, C^2 = -1$).
The generators of the spinor representations  are $S^{MN} = -\frac12 \sigma^{MN}\,$, with
\be
(\sigma^{MN})^a_{\,\,b} =\frac12(\tilde{\gamma}^M\gamma^N-
\tilde{\gamma}^N\gamma^M)^a_{\,\,b} = \frac12({\gamma}^N\tilde{\gamma}^M-
{\gamma}^M\tilde{\gamma}^N)^{\,\,a}_{b}\,, \quad M,N = 0, \ldots 5\,,
\ee
where gamma matrices $(\gamma^M)_{ac}$  and $(\tilde{\gamma}^N)^{cb}$ are antisymmetric $4\times 4$ matrices satisfying the $6D$ Clifford algebra relation
\be
(\gamma^M)_{ac}(\tilde\gamma^N)^{cb}+(\gamma^N)_{ac}(\tilde\gamma^M)^{cb}
=-2\delta^b_a \eta^{MN}\,, \quad \eta^{MN} = {\rm diag}(1, -1, -1, -1, -1, -1),
\ee
and
\be
(\tilde{\gamma}_M)^{ab}=\frac12\varepsilon^{abcd} (\gamma_M)_{cd}\,, \quad \varepsilon_{1234} = \varepsilon^{1234} = 1\,.
\ee
Some other useful properties of gamma matrices (employed in Sect. 7) are
\bea
&& (\gamma_M)_{ab}\,(\gamma_M)_{cd} = 2\varepsilon_{abcd}\,, \quad (\tilde{\gamma}_M)^{ab}\,(\tilde{\gamma}_M)^{cd} = 2\varepsilon^{abcd}\,, \nn
&& (\sigma^{MN})^a_b\, (\sigma_{MN})^c_d = 2\big(\delta^a_b\,\delta^c_d - 4 \delta^a_d\,\delta^c_b\big), \nn
&& {\rm Tr} (\sigma^{MN}\, \sigma^{ST}) = 4\big(\eta^{NS}\,\eta^{MT} -  \eta^{MS}\,\eta^{NT} \big). \lb{GammaRel}
\eea

In the spinor notation, $6D$ vectors are uniquely represented as
\be
V_M \; \leftrightarrow \; V_{ab} := \frac12 (\gamma^M)_{ab} V_M\,, \quad V_M = \frac12 \tilde{\gamma}_M^{ab}V_{ab}\,. \lb{VecSpin}
\ee
Respectively, $6D$ Minkowski space can be parameterized by the coordinates
$x^{ab}$ related to the standard $6D$ coordinates $X^M, M = 0, \ldots, 5$ as
\be
x^{ab} = (\tilde{\gamma}^M)^{ab}x_M\,, \quad \partial_{ab} = \frac12 (\gamma^M)_{ab}\partial_M\,. \lb{Mab}
\ee
In what follows, we shall make use of the superspace approach in which $6D$ Minkowski space is extended by
anticommuting Grassmann odd spinor coordinates $\theta^a_i$ and $\theta_{a A}$, with $i = 1, 2; A = 1, 2 $ being doublet indices of
some external automorphism ($R$-symmetry) $SU(2)$ groups.

The standard superspace approach is not too useful for gauge theories  with extended supersymmetry.
First, it does not allow to reveal the basic unconstrained objects of the theory and to construct the relevant off-shell superfield actions. Secondly,
the standard superspace constraints on the basic gauge superfields frequently put the theory on mass shell and give no simple hints as to which portion of
the total supersymmetry still admits an off-shell realization. Much more suggestive is the harmonic superspace (HSS) approach firstly discovered for $4D\,, {\cal N}=4$
supersymmetry in \cite{20,18}. Off-shell superfield formulations of both $6D, {\cal N} =(1,0)$ SYM theory and hypermultiplets become possible
within $6D, {\cal N} =(1,0)$ HSS \cite{19,22}, a direct generalization of $4D, {\cal N}=2$ HSS. On the other hand, in an analogous formulation of the maximally extended
${\cal N} =(1,1)$ SYM theory in $6D$ HSS with the double set of harmonic and Grassmann variables \cite{25,12} the total ${\cal N}= (1, 1)$ Poincar\'e supersymmetry
admits a closure on the corresponding harmonic superfields only on shell. Hence,  the maximally supersymmetric ${\cal N}= (1, 1)$ invariants can also be defined on shell
(though they can still respect off-shell ${\cal N} =(1,0)$ supersymmetry).

Now we sketch the basic notions of ${\cal N}= (1, 0)$ HSS. The ${\cal N}= (1, 0)$ super Poincar\'e group in its bosonic sector involves the $SU(2)$
automorphisms ($R$-symmetry) group, besides $6D$ Poincar\'e group. When defining the
standard ${\cal N}= (1, 0)$ superspace, this $SU(2)$ is factored out, so this superspace
is parameterized by the coordinate set
\be
z := (x^{ab}, \, \theta^c_i)\,.\lb{10SS}
\ee
${\cal N}=(1,0)$ HSS is obtained by factoring out only the $U(1)$ part of the automorphism  $SU(2)$. Thus each superspace point is
equipped with the additional harmonic variables $u^{\pm i}$ which are the coordinates of the sphere $S^2\simeq SU(2)/U(1)\,$, $u^{+ i}u^-_i = 1$. This extended coordinate
set yields the {\it central} basis of ${\cal N}= (1, 0)$ HSS
\be
Z := (z, u) = (x^{ab}, \theta^c_i, u^{\pm i})\,.\lb{10HSS}
\ee

It is convenient to define an {\it analytic} basis in this HSS, by passing to the coordinates
\be
Z^{(an)} := (x^{(an)(ab)}, \theta^{\pm c}, u^{\pm i})\,,\lb{10HSSa}
\ee
where
\be
x^{(an) ab} =x^{ab} + \frac i2 (\theta^{+a}\theta^{-b} - \theta^{+b}\theta^{-a}),\quad \theta^{\pm a}=
u^\pm_k\theta^{ak}.
\ee
Now ${\cal N}=(1,0)$ supersymmetry transformations prove to be closed also on the reduced set of coordinates
 \be
\zeta :=(x^{(an)ab}, \theta^{+c}, u^{\pm i}) \,, \label{AnalSS}
\ee
that parametrize a Grassmann-analytic subspace \footnote{Both coordinate sets \p{10HSSa} and \p{AnalSS} are real under the generalized conjugation defined in \cite{18},
so the latter is indeed a subspace of the former.}. Since G-analytic superfields depend only on
the half of odd variables, their component structure is simpler than that of general harmonic superfields. In
order to define the G-analyticity condition, we introduce the spinor differential operators
\be
D^+_a=\partial_{-a}~, \quad D^-_a=-\partial_{+ a}-2i\theta^{-b}\partial_{ab}\,, \quad \{D^+_a,D^-_b\}=2i\partial_{ab}\,,
\ee
which are related to the spinor derivatives in the central basis \p{10HSS} by the standard relations, $D^\pm_a= D_a^iu^\pm_i$.
Then an arbitrary G-analytic superfield $\Phi(\zeta)$ can be defined as the one satisfying the Grassmann analyticity condition
\be
D^+_a \Phi = 0\,.
\ee
Like in $4D$ case \cite{18}, the existence of the analytic harmonic superspace with the half set of Grassmann coordinates and the opportunity
to define ``short'' harmonic analytic superfields are the pivotal merits of the $6D$, ${\cal N}=(1,0)$ HSS approach.

The presence of additional bosonic coordinates brings about new algebraic
relations, those involving the harmonic derivatives,
\be
[D^{++}, D^{--}] = D^0\,,
\ee
where (in the analytic basis)
\bea
&&D^{\pm\pm} =\partial^{\pm\pm}+i\theta^{+a}\theta^{+b}\partial_{ab}+\theta^{+a}\partial_{-a}~,\quad \partial^{\pm\pm} =
u^{\pm i} \frac {\partial }{ \partial u^{\mp i}}\ , \nn
&& D^0 = u^{+i} \frac {\partial}{ \partial u^{+i}} -
u^{-i} \frac {\partial}{ \partial u^{-i}} + \theta^{+a} \partial_{+ a} -  \theta^{-a} \partial_{- a}\,.
\eea
The commutation relations between harmonic and spinor derivatives can be easily established
\be
[D^{++},D^+_a]=[D^{--},D^-_a]=0\, ,\quad [D^{++},D^-_a]=D^+_a,\quad [D^{--},D^+_a]=D^-_a.
\ee

As was said earlier, an on-shell superfield description of $6D, {\cal N}=(1,1)$ SYM theory is achieved in the framework of bi-harmonic ${\cal N}=(1,1)$ superspace.
It is an extension of the superspace \p{10HSS} by new right-handed $Spin(5,1)$ spinors $\theta_{a A}$, where on the doublet indices $A= 1,2$
 an additional $R$-symmetry $SU(2)$ group acts, and by the corresponding extra set of harmonics $u^{\hat{\pm}}_A\,, \; u^{\hat{+} A} u^{\hat{-}}_ A = 1$,
\bea
&& Z \;\Rightarrow \; \hat{Z} := (x^{(an)ab}\,, \,\theta^{\pm a}, \, \theta^{\hat{\pm}}_a, u^{\pm}_i, \,u^{\hat{\pm}}_A)\,, \nn
&& \theta^{\hat{\pm}}_a = \theta^{A}_a u^{\hat{\pm}}_A, \,\quad
x^{(an)ab} = x^{ab} +\frac i2 (\theta^{+a}\theta^{-b} - \theta^{+b}\theta^{-a}) + \frac i2 \varepsilon^{abcd}\theta^{\hat{+}}_c\theta^{\hat{-}}_d\,. \lb{hatZ}
\eea
In this basis, both spinor derivatives $D^{+}_a$ and $D^{\hat{+}a}$ are short,
\bea
D^{+}_a = \frac{\partial}{\partial \theta^{-a}}\,, \quad D^{\hat{+}a} = \frac{\partial}{\partial \theta^{\hat{-}}_{a}}\,,
\eea
which reflects the existence of the closed analytic subspace with the double set of Grassmann and harmonic coordinates:
\be
\hat\zeta := (x^{(an)ab}\,, \,\theta^{+ a}, \, \theta^{\hat{+}}_a, u^{\pm i}, \,u^{\hat{\pm}}_A)\,.\lb{AnalSS2}
\ee

Note that one can impose the Grassmann analyticity conditions either separately with respect to $\theta^{- a}$ and $\theta^{\hat{-}}_a$, or simultaneously
with respect to both these coordinates, without contradiction with the coordinate action of ${\cal N}=4$ Poincar\'e supersymmetry in the basis \p{hatZ}.
The second option corresponds to the minimal-dimension ``$1/2$'' analytic subspace \p{AnalSS2}, while the first one  to the existence of the additional ``$3/4$'' analytic subspaces
$\hat{\zeta}_I$ and $\hat{\zeta}_{II}$ parametrized by the following sets of coordinates
\be
\hat{\zeta}_I = (\hat{\zeta}, \theta^{-a})\,, \quad \hat{\zeta}_{II} = (\hat{\zeta}, \theta^{\hat{-}}_a)\,. \lb{IandII}
\ee
They are also closed under total ${\cal N}=(1,1)$ supersymmetry. While $\hat{\zeta}$ involves eight Grassmann coordinates compared to sixteen such coordinates
 of the full ${\cal N}=(1,1)$ HSS, both $3/4$ analytic subspaces contain twelve Grassmann coordinates.

The harmonic derivatives with respect to $u^{\hat{\pm}}_A$ in the analytic basis have the same structure as the ${\cal N}=(1,0)$ ones defined earlier, modulo
the proper altering of the position of the spinor indices. The same concerns their mutual commutators, their commutation relations with the hat-spinor derivatives, and the
mutual anticommutators of the hat-spinor derivatives, taking into account the relation
$$
\partial^{ab} = \frac12 \varepsilon^{abcd}\partial_{cd}\,.
$$

To complete this section, we define the integration measures for
the full and analytic ${\cal N} = (1,0)$ and${\cal N} = (1,1)$ HSS
\bea
&&d {Z}=d^6x^{(an)} du \,(D^-)^4(D^+)^4,\quad  d\zeta^{(-4)}=d^6x^{(an)} du (D^-)^4\,, \lb{10Integ} \\
&& d \hat{Z} = dZ d\hat{u} (D^{\hat{-}})^4(D^{\hat{+}})^4\,, \quad d\hat{\zeta}^{(-4, \hat{-}4)}= d\zeta^{-4}d\hat{u}(D^{\hat{-}})^4\,, \lb{11Int-a}\\
&& d \hat{\zeta}^{(0,\hat{-}4)}_{I} = dZ d\hat{u}(D^{\hat{-}})^4\,, \quad d \hat{\zeta}^{({-}4, 0)}_{II} = d\zeta^{(-4)} d\hat{u}(D^{\hat{-}})^4(D^{\hat{+}})^4\,, \lb{11Int-b} \\
&& (D^\pm)^4=-\frac{1}{24}\varepsilon^{abcd}D^\pm_a D^\pm_b D^\pm_c D^\pm_d\,, \quad (D^{\hat{\pm}})^4=-\frac{1}{24}
\varepsilon_{abcd}D^{\hat{\pm}a} D^{\hat{\pm}b} D^{\hat{\pm}c} D^{\hat{\pm}d}\,. \lb{DefD4}
\eea

\setcounter{equation}{0}

\section{${\cal N} = (1, 0)$ supersymmetric Yang-Mills theory}

\subsection{Kinematics}
Now we turn to supersymmetric Yang-Mills (SYM) theory in $6D, {\cal N}=(1,0)$ HSS \cite{19,22,17,12}\footnote{See also \cite{24,Kuz1,Kuz2}.}.  First, we
introduce the gauge superfield $V^{++}$ that covariantizes the harmonic derivative
\be
\nabla^{++} = D^{++} + V^{++}\,. \lb{DV++}
\ee
The connection $V^{++}$ is a G-analytic superfield that transforms as $\delta V^{++} = \nabla^{++}\Lambda$
under gauge transformations with a G-analytic gauge parameter $\Lambda(\zeta)$. The  second harmonic
derivative $D^{--}$ is covariantized by a non-analytic connection $V^{--}$,
\be
\nabla^{--} = D^{--} + V^{--}\,.\lb{DV--}
\ee
This connection is uniquely determined from the zero-curvature (``harmonic flatness'')
condition,
\be
[\nabla^{++}, \nabla^{--}] = D^0 \; \Rightarrow \; D^{++}V^{--} - D^{--}V^{++} + [V^{++}, V^{--}] =0\,, \lb{Flat10}
\ee
as a series over the connections $V^{++}$ given at different harmonic ``points'':
\be
\label{Vmmrjad}
V^{--}(z,u)=\sum\limits^{\infty}_{n=1} (-1)^n \int du_1\ldots du_n\,
\frac{V^{++}(z,u_1 )\ldots V^{++}(z,u_n )}{(u^+ u^+_1)(u^+_1 u^+_2)\ldots (u^+_n u^+)}\,,
\ee
where $1/(u^+u^+_1), \ldots $ are harmonic distributions defined in \cite{18}. It is straightforward
to show  that $\delta V^{--} = \nabla^{--}\Lambda$ \footnote{This and many other properties can be proved using the Lemma: $\nabla^{++} F^{-n} = 0 \;\Rightarrow \,
F^{-n} =0$ for $n\geq 1$, where $F^{-n}$ is a harmonic ${\cal N}=(1,0)$ superfield.} .

Defining
\be
{\cal D}^-_a = D^{- a} + {\cal A}^-_a\,, \quad {\cal D}_{ab} = \partial_{ab} + {\cal A}_{ab}
\ee
and imposing the commutation relations
\bea
&& [\nabla^{--}, D^+_a] = {\cal D}^{-}_a\,, \quad [\nabla^{++}, {\cal D}^{-}_a] = D^+_a\,, \quad [\nabla^{++}, D^{+}_a]
= [\nabla^{--}, {\cal D}^{-}_a] = 0\,, \nn
&& \{D^+_a, {\cal D}^-_b\} = 2i {\cal D}_{ab}\,,
\eea
we can express the spinor and vector connections in terms of $V^{--}$:
\be
{\cal A}^-_a(V)=-D^+_a V^{--},\qquad {\cal A}_{ab}(V)= \frac{i}{2}D^+_aD^+_b V^{--}\,. \lb{A-Aab}
\ee

The covariant spinor superfield strength $W^{+ a}$ is defined as
\be
[D^+_a, {\cal D}_{bc}] \ =\ \frac{i}2\varepsilon_{abcd} W^{+d} \, , \quad W^{+a}=-\frac{1}{6}\varepsilon^{abcd}D^+_b D^+_c D^+_d V^{--}\,.\lb{W+Def}
 \ee

For later use, we also define
\be
W^{-a} := \nabla^{--} W^{+ a}\,, \quad \delta W^{\pm a} = -[\Lambda, W^{\pm a}]\,, \lb{W-def}
\ee
so that
\be
[{\cal D}^-_a, {\cal D}_{bc}] \ =\ \frac{i}2\varepsilon_{abcd} W^{-d}\,.\lb{W-DefId}
\ee
The spinor superfield strengths obey the following off-shell relations
\bea
&& \nabla^{++} W^{+ a} =\nabla^{--} W^{- a} = 0\,, \quad \nabla^{++} W^{- a} = W^{+a}\,, \nn
&& D^{+}_b W^{+ a} = \delta^a_b F^{++}\,, \quad F^{++} :=(D^+)^4 V^{--} = \frac14 D^+_a W^{+ a}\,, \lb{Wident1} \\
&& \nabla^{++}F^{++} = D^+_a F^{++} = 0 \lb{Wident5}
\eea
(as well as some other ones derived through acting by $\nabla^{--}$ on eqs. \p{Wident1}, \p{Wident5}). An important Bianchi identity
stemming  from \p{W+Def} and \p{W-DefId} is
\bea
[{\cal D}_{ab}, {\cal D}_{cd}] = \frac14 \big(\varepsilon_{acdf}{\cal D}^-_b W^{+ f} + \varepsilon_{bcdf}{D}^+_a W^{- f}\big), \lb{nabla-nabla}
\eea
which, in particular, implies
\bea
&&{\cal D}^-_b W^{+ b} =  D^+_b W^{-b}\,, \lb{nabla-nabla1} \\
&& [{\cal D}^{ab}, {\cal D}_{cb}] = \frac14\big( D^+_c W^{- a} - {\cal D}^-_c W^{+ a} \big)\,, \lb{nabla-nabla2} \\
&& {\cal D}_{ab}W^{+ b} = \frac{i}8 {\cal D}^{-}_a D^+_c W^{+ c}\,, \quad {\cal D}_{ab}W^{- b} = -\frac{i}8 D^{+}_a {\cal D}^-_c W^{-c}\,. \lb{nablaW5}
\eea

\subsection{Dynamics}
The superfield action of  $6D, {\cal N}=(1,0)$ SYM theory can be written
in the form \cite{22}
   \be
S^{SYM} =\frac{1}{f^2}\sum\limits^{\infty}_{n=2} \frac{(-1)^{n}}{n} {\rm Tr} \int
d^6\!x\, d^8\theta\, du_1\ldots du_n \frac{V^{++}(z,u_1 )
\ldots V^{++}(z,u_n ) }{(u^+_1 u^+_2)\ldots (u^+_n u^+_1 )}\,,\lb{action1}
   \ee
where $f$ is a coupling  constant  with the dimension of inverse mass, so that the relevant component Lagrangian density
has the canonical (``engineering'') dimension $d= 4$. The action \p{action1} is invariant under
the supergauge transformations of $V^{++}$ and yields the equation of motion
\be
F^{++} = (D^+)^4 V^{--} = 0\,.\lb{EqMV}
\ee
Both these properties can be derived by employing the variation formula
\be
\delta S^{SYM} = -\frac{1}{f^2} {\rm Tr} \int dZ \delta V^{++} V^{--} = -\frac{1}{f^2} {\rm Tr} \int \zeta^{(-4)} \delta V^{++} F^{++}\,.\lb{VarV}
\ee

Using the representation
\be
F^{++} = \frac14 D^+_a W^{+ a}\,,
\ee
eq. \p{EqMV} can be rewritten in terms of the spinor superfield strength as
\be
D^+_a W^{+ a} = 0\, \quad {\rm or} \quad  D^+_a W^{+ b} = 0\,. \lb{EqmW}
\ee

Due to the gauge freedom with the analytic super-parameter $\Lambda$, one can choose the Wess-Zumino gauge for $V^{++}$
\bea
&& V^{++} = i\theta^{+a}\theta^{+ b}A_{ab} + \frac13 \Psi^{+ 3}_a \lambda^{-a} + \frac18 \Psi^{+4}{\cal D}^{-2}, \nn
&&\lambda^{-a} = \lambda^{ia}(x) u^-_i\,, \quad {\cal D}^{-2} = {\cal D}^{(ik)}(x)u^-_iu^-_k\,,\lb{WZ}
\eea
where
\be
\Psi^{+ 3}_d := \varepsilon_{abcd}\theta^{+a}\theta^{+b}\theta^{+c}\,, \quad \Psi^{+4} := \varepsilon_{abcd}\theta^{+a}\theta^{+b}\theta^{+c}\theta^{+d}\,.
\ee
The fields entering \p{WZ} form the component off-shell $6D, {\cal N}=(1,0)$ SYM multiplet.

Along with $V^{++}$, an important ingredient of the harmonic superspace formulation of $6D, {\cal N}=(1,1)$ SYM theory is the analytic
hypermultiplet superfield $q^{+ A}(\zeta)$ which also takes values in the adjoint representation of the gauge group and satisfies the reality condition
\be
\widetilde{q^{+ A}} = \varepsilon_{AB} q^{+ B}\,.
\ee
Its action reads
\be
S^q = \frac{1}{2f^2}\,{\rm Tr} \int d\zeta^{(-4)} q^{+ A}\nabla^{++} q^+_A\,.
\ee
The sum of two actions,
\be
S_{(1,1)} = S^{SYM} + S_q, \lb{Act11}
\ee
reveals a hidden ${\cal N} = (0,1)$ supersymmetry \cite{12},
\be
\delta_{(0,1)}V^{++} = \epsilon^{+ A}q^+_A\,, \quad \delta_{(0,1)}q^{+ A} = -(D^+)^4 (\epsilon^{-A}V^{--})\,, \;\; \epsilon^{\pm A} :=\epsilon_a^A\theta^{\pm a}\,,\lb{11Susy}
\ee
where $\epsilon_a^A$ is the corresponding Grassmann parameter \footnote{The transformations \p{11Susy} provide an {\it off-shell} invariance of \p{Act11}. However,
their closure with the manifest ${\cal N}=(1,0)$ supersymmetry amounts to the standard relations of $6D\,, {\cal N}=(1,1)$ Poincar\'e superalgebra
only {\it on shell}.}.  So the total action describes ${\cal N} = (1,1)$ SYM theory. The equation
of motion for the hypermultiplet is
\be
\nabla^{++}q^{+ A} = 0\,. \lb{EqMq}
\ee
The equation of motion for $V^{++}$ following from the action $S_{(1,1)}$ undergoes the evident modification as compared to \p{EqMV}:
\be
F^{++} + \frac12 [q^{+A}, q^+_A] = 0\,. \lb{EqMV+q}
\ee

On the mass shell, the previously listed Bianchi identities are simplified and some new relations of this kind can be derived. In the limit
of vanishing hypermultiplet superfield (this limit will be employed in Sect. 7), the equations of motion \p{EqmW}, combined with the relation \p{nabla-nabla1},
imply
\bea
D^{+}_a W^{- b} = - {\cal D}^{-}_a W^{+ b}\,, \quad D^{+}_a W^{- a} = {\cal D}^{-}_a W^{+ a} = {\cal D}^{-}_a W^{- a} = 0\,.\lb{EquivEm}
\eea
The Bianchi identity \p{nablaW5} yields
\bea
{\cal D}_{ab} W^{\pm b} = 0\,, \lb{cyclic1}
\eea
or, in an equivalent form,
\bea
{\cal D}^{ab} W^{\pm c} + {\cal D}^{ca} W^{\pm b} + {\cal D}^{bc} W^{\pm a} = 0\,. \lb{cyclic}
\eea
The last two identities, together with \p{nabla-nabla2}, give rise to the equation
\be
{\cal D}_{ab}{\cal D}^{ab} W^{\pm c} = -[D^+_b W^{- c}, W^{\pm b}] = [{\cal D}^-_b W^{+ c}, W^{\pm b}]\,. \lb{Dsquare}
\ee
Eqs. \p{cyclic1} and \p{Dsquare} are none other than the covariantized $6D$ Dirac and Klein-Fock-Gordon equations for the ${\cal N}=(1,0)$ vector multiplet.
In the abelian limit they are reduced to
\be
\partial_{ab} W^{\pm b} = 0\,, \quad \partial_{ab}\partial^{ab} W^{\pm c} = 0\,. \lb{DirDea}
\ee

\section{${\cal N} = (1,1)$ SYM in bi-HSS formulation}
When $6D, {\cal N}=(1,1)$ SYM theory is formulated in terms of ${\cal N}=(1, 1)$ superfields
(\cite{16}, \cite{11} and \cite{25}), the constraints of the theory imply equations of
motions, {\it i.e.} this formulation supplies an example of  the {\it on-shell} description.

The original ${\cal N}=(1,1)$ SYM constraints, before introducing the harmonic coordinates,
are written in terms of the gauge-covariantized spinor derivatives,
  \be
\lb{spincovder}
\nabla^i_a \ =\ \frac \partial {\partial \theta^a_i} - i \theta^{bi} \partial_{ab} + {\cal A}^i_a \, ,\quad
\hat{\nabla}^{aA} \ =\ \frac \partial {\partial \hat{\theta}_{Aa}} - i \hat{\theta}_b^A \partial^{ab}  +
\hat{{\cal A}}^{aA} \, ,
  \ee
where ${\cal A}^i_a$ and $ \hat{{\cal A}}^{aA}$
 are the spinor connections and the convention
 $\nabla^{ab} = \tfrac{1}{2} \varepsilon^{abcd}
 \nabla_{cd}$ is assumed. The constraints read \cite{16,11}
\be
\lb{spincomm0}
\{\nabla^{(i}_a, \nabla^{j)}_b \} \ =\ \{\hat{\nabla}^{a(A}, \hat{\nabla}^{bB)} \} = 0 \, ,
 \ee
 \be
\lb{spincommphi}
\{\nabla^i_a, \hat{\nabla}^{bA} \} \ =\ \delta_a^b \phi^{iA} \, .
 \ee
Through the Bianchi identities, eqs. \p{spincomm0} and \p{spincommphi} imply
 \be
\lb{const-phiiA}
 \nabla^{(i}_a \phi^{j)A}  \ =\   \hat{\nabla}^{a(A}   \phi^{iB)} \ =\ 0 \, .
 \ee
The  constraints \p{spincomm0} and \p{spincommphi}
 define $6D, {\cal N}=(1,1)$ SYM supersymmetric theory in the standard  ${\cal N}=(1,1)$ superspace parametrized by the coordinates $(x^{ab},\theta^a_i, \hat{\theta}_{Aa})$.
As was already mentioned, they imply the equations of motion for the  superfields involved and so put the theory on shell.

The standard steps of the harmonic interpretation of the above constraints can be summarized as: ({\bf i}) contracting the $SU(2)$ indices of the spinor derivatives
and superfield strengths with the harmonics $u^{\pm}_i,  u^{\hat{\pm}}_A$,
thereby passing to the harmonic projections of these derivatives; ({\bf ii}) replacing the property of linearity in harmonics by the property of vanishing of some commutation relations
with the harmonic derivatives; ({\bf iii}) passing to the analytic basis and frame, so as to make ``short'' ({\it i.e.} having no connection in this frame) as much spinor derivatives as possible.
After these steps  the constraints are rewritten as the integrability conditions for some sort of Grassmann analyticity, while the underlying gauge multiplet proves to be carried by the harmonic
gauge connections covariantizing flat harmonic derivatives.

This general scheme, being applied to the case at hand, yields the following equivalent set of the constraints \cite{12}
\bea
&& \{\nabla_a^+, \nabla_b^+\} = \{D^{\hat{+} a} , D^{\hat{+} b}\} = 0\,, \quad \{\nabla_a^+, D^{\hat{+}b}\} = \delta^b_a\,\phi^{+\hat{+}}\,, \nn
&& [\nabla^{++}, \nabla^+_a] = [\nabla^{\hat{+}\hat{+}}, \nabla^+_a] = [\nabla^{++}, D^{\hat{+}a}] = [\nabla^{\hat{+}\hat{+}}, D^{\hat{+}a}] = 0\,, \nn
&&[\nabla^{++}, \nabla^{\hat{+}\hat{+}}] = 0\,, \lb{ConstrHarm}
\eea
where we have chosen the ``hat-analytic'' basis, with a short hat-spinor derivative, $\nabla^{\hat{+} a} = D^{\hat{+} a} = \partial/\partial\theta^{\hat{-}}_a$.
All other derivatives have non-trivial gauge connections
\bea
\nabla^+_a = D^+_a + {\cal A}^+_a\,, \quad \nabla^{++} = D^{++} + \tilde{V}^{++}\,, \quad \nabla^{\hat{+}\hat{+}} = D^{\hat{+}\hat{+}} + V^{\hat{+}\hat{+}}\,.\lb{11Connections}
\eea
The constraints \p{ConstrHarm} are covariant under the gauge transformations with the $3/4$ hat-analytic parameter $\tilde{\Lambda}$:
\bea
&& \delta \tilde{V}^{++} = \nabla^{++}\tilde{\Lambda}\,, \;\delta {V}^{\hat{+}\hat{+}} = \nabla^{\hat{+}\hat{+}}\tilde{\Lambda}\,, \; \delta \mathcal{A}^+_a = \nabla^+_a \tilde{\Lambda}\,, \quad \delta \phi^{+\hat{+}} = [\phi^{+\hat{+}}, \tilde{\Lambda}] \nn
&& D^{\hat{+}a}\tilde{\Lambda} = 0 \; \Rightarrow \; \tilde{\Lambda} = \tilde{\Lambda}(\hat{\zeta}, \theta^{- a}) = \tilde{\Lambda}(\hat{\zeta}_I)\,.
\eea

Eqs. \p{ConstrHarm} impose non-trivial on-shell conditions on the gauge connections. The notable property of the constraints in the form
\p{ConstrHarm} is that the derivative $\nabla^+_a$ cannot be chosen ``flat'' simultaneously with  $\nabla^{\hat{+}a}$ because of their non-vanishing
anticommutator.

The full set of the relations solving \p{ConstrHarm} in terms of ${\cal N}= (1,0)$ superfields was found in \cite{12}. Below we shall briefly summarize them.

\begin{itemize}
\item An important result is
that, after imposing the appropriate gauge fixing and due to the vanishing of the commutator between the hatted and unhatted
covariant harmonic derivatives, the connections $V^{++}$  and $V^{--}$ prove to coincide with their ${\cal N} =(1, 0)$ prototypes,
\be
\tilde{V}^{++} = V^{++}(\zeta)\,, \quad D^{++}V^{--} - D^{--}V^{++} + [V^{++}, V^{--}] = 0\,.
\ee

\item For the ${\cal N}=(1,1)$ SYM superfield strength $\phi^{+\hat{+}}$ the following solution was obtained
\bea
\phi^{+\hat{+}} &=& q^{+ \hat{+}} - \theta^{\hat{+}}_a W^{+ a} -
i \theta^{\hat{+}}_a\theta^{\hat{+}}_b{\cal D}^{ab}q^{+\hat{-}}
+\frac16 \Psi^{\hat{+}3 d}[D^+_d q^{-\hat{-}}, q^{+ \hat{-}}]  \nn
&&+\, \frac1{24} \Psi^{\hat{+}4} [q^{+\hat{-}}, [q^{+\hat{-}}, q^{-\hat{-}}]]\,,\lb{explphi}
\eea
where $\Psi^{\hat{+}3 d} := \varepsilon^{abcd}\theta^{\hat{+}}_{a}\theta^{\hat{+}}_{b}\theta^{\hat{+}}_{c}\,,
\; \Psi^{\hat{+}4} := \varepsilon^{abcd}\theta^{\hat{+}}_{a}\theta^{\hat{+}}_{b}\theta^{\hat{+}}_{c}\theta^{\hat{+}}_{d}\,,$ $W^{+ a}$ was defined in \p{W+Def},
$q^{\pm\hat{\pm}} = q^{\pm A}u^{\hat{\pm}}_A\,, \; q^{\pm\hat{\mp}} = q^{\pm A}u^{\hat{\mp}}_A\,,$ and
$$
q^{-A}:= \nabla^{--} q^{+ A}\,.
$$
Recall that ${\cal D}^{ab} = \frac12 \varepsilon^{abcd} {\cal D}_{cd}\,, \quad {\cal D}_{cd} = \partial_{cd} + {\cal A}_{cd}$
and the vector ${\cal N}=(1,0)$ connection ${\cal A}_{cd}$ was defined in \p{A-Aab}.

\item The final expression for the harmonic connection $V^{\hat{+}\hat{+}}$  is as follows
\bea
V^{\hat{+}\hat{+}} =i \theta^{\hat{+}}_a\theta^{\hat{+}}_b {\cal A}^{ab} -\frac13 \Psi^{\hat{+}3 a} D^+_a q^{-\hat{-}} + \frac18\Psi^{\hat{+}4}[q^{+\hat{-}}, q^{-\hat{-}}]\,.\lb{HatV++}
\eea
As a consequence of the second flatness condition,
\be
[\nabla^{\hat{+}\hat{+}}, \nabla^{\hat{-}\hat{-}}] = \hat{D}^0 \; \Leftrightarrow \; D^{\hat{+}{+}}V^{\hat{-}\hat{-}} - D^{\hat{-}\hat{-}}V^{\hat{+}\hat{+}}
+ [V^{\hat{+}\hat{+}}, V^{\hat{-}\hat{-}}] = 0\,, \lb{2ndflatness}
\ee
the non-analytic harmonic potential $V^{\hat{-}\hat{-}}$ entering $\nabla^{\hat{-}\hat{-}} = D^{\hat{-}\hat{-}} + V^{\hat{-}\hat{-}}$
can be expressed through $V^{\hat{+}\hat{+}}$.

\item The constraints \p{ConstrHarm} accompanied by the suitable gauge choice fix the form of the spinor connection ${\cal A}^+_a$ entering $\nabla^+_a$ in \p{11Connections} as
\bea
{\cal A}^+_a = -\theta^{\hat{+}}_a q^{+\hat{-}} + \theta^{\hat{-}}_a \phi^{+\hat{+}}\,.\lb{A+}
\eea

\item The constraints also imply the equations of motion \p{EqMV+q}, \p{EqMq} for the superfields $V^{++}$ and $q^{+ A}$. Using these equations of motion together
with the explicit formulas quoted above (as well as some remote consequences of them collected in Appendix A of \cite{12}), one can check the important properties
\be
\nabla^{++}\phi^{+\hat{+}} = \nabla^{\hat{+}\hat{+}}\phi^{+\hat{+}} = 0\,,\lb{HarmConstrphi}
\ee
which, in fact, directly follow from the constraints \p{ConstrHarm} as the Bianchi identities. For further use, we also define
\be
\phi^{-\hat{+}} := \nabla^{--}\phi^{+\hat{+}}\,, \;\;  \phi^{+\hat{-}} := \nabla^{\hat{-}\hat{-}}\phi^{+\hat{+}}\,, \;\;  \phi^{-\hat{-}} =
\nabla^{\hat{-}\hat{-}}\nabla^{--}\phi^{+\hat{+}}\,. \lb{DefProjphi}
\ee
It is easy to check that
\be
\nabla^{--}\phi^{-\hat{\pm}} =\nabla^{++}\phi^{+\hat{-}} = \nabla^{\hat{-}\hat{-}}\phi^{\pm \hat{-}} = \nabla^{\hat{+}\hat{+}}\phi^{-\hat{+}} = 0\,, \lb{Analphi2}
\ee
and
\be
\nabla^{++}\phi^{-\hat{\pm}} = \phi^{+\hat{\pm}}\,, \;\;\nabla^{\hat{+}\hat{+}}\phi^{\pm\hat{-}} = \phi^{\pm\hat{+}}\,.\lb{Relphi}
\ee

\item
The superfield strength $\phi^{+\hat{+}}$ also satisfies the on-shell analyticity conditions
\be
\nabla^{+}_a \phi^{+\hat{+}} = D^{\hat{+}a}\phi^{+\hat{+}} = 0\,, \lb{twoAnal}
\ee
which follow as the Bianchi identities from \p{ConstrHarm} and can be explicitly checked using the relations given above,
the ${\cal N}=(1,0)$  analyticity of $q^{+A}$, $D^+_a q^{+A} =0$, and the ${\cal N}=(1,0)$ superfield equations of motion.

\item
After all gauge-fixings and solving the constraints, the only residual gauge freedom is the ${\cal N}=(1,0)$ gauge freedom with the analytic
${\cal N}=(1,0)$ gauge parameter, $\tilde{\Lambda}(\hat{\zeta}, \theta^-_a) \, \rightarrow\, \Lambda(\zeta)\,$.

\end{itemize}

As the last steps in constructing the formalism of the differential superspace geometry of ${\cal N}=(1,1)$ SYM theory, one defines, in the standard way,
the negatively charged covariant spinor derivatives
\bea
\nabla^-_a := [ \nabla^{--}, \nabla^+_a]\,, \quad   \nabla^{\hat{-}a} := [ \nabla^{\hat{-}\hat{-}} , D^{\hat{+}a}] \lb{-nabla}
\eea
and the covariant vector derivatives
\bea
&& \{D^{\hat{+}a},\nabla^{\hat{-}b} \} := 2i\nabla^{ab} = 2i(\partial^{ab} +{\cal V}^{ab})\,, \quad  {\cal V}^{ab} = \frac{i}{2} D^{\hat{+}a}D^{\hat{+}b}V^{\hat{-}\hat{-}}\,,\lb{Vect1} \\
&& \{\nabla^{{+}}_{a},\nabla^{{-}}_{b} \} := 2i\nabla_{ab} = 2i(\partial_{ab} +{\cal V}_{ab})\,, \quad {\cal V}_{ab} = \frac12\varepsilon_{abcd}{\cal V}^{cd}\,.  \lb{Vect2}
\eea
Note that the identification of the vector connections obtained in two different ways plays an important role in finding the explicit solution of the constraints
\p{ConstrHarm} presented above. Also note that, taking into account the expression \p{A+},  the spinor derivative $\nabla^-_a$ defined in \p{-nabla} has the very simple form
\be
\nabla^-_a = {\cal D}^-_a - \theta^{\hat{+}}_a q^{-\hat{-}} + \theta^{\hat{-}}_a \phi^{-\hat{+}}\,.\lb{-nabla2}
\ee

In what follows, we will also use the relations
\bea
&& \{ D^{\hat{+}a}, \nabla^{-}_b\} = \delta^a_b \phi^{-\hat{+}}\,, \quad  \{ \nabla^{\hat{-}a}, \nabla^{\pm}_b\} = \delta^a_b \phi^{\pm\hat{-}}\,, \lb{PolRela}\\
&& D^{\hat{+}a}\phi^{-\hat{-}} = -\nabla^{\hat{-}a}\phi^{-\hat{+}}\,, \; \nabla^{+}_{a}\phi^{-\hat{+}} =
-\nabla^{-}_a \phi^{+\hat{+}}\,, \; \nabla^{-}_a \phi^{-\hat{\pm}} = \nabla^{\hat{-}a}\phi^{\pm\hat{-}} = 0\,, \lb{PolRelb}\\
&& D^{\hat{+}a}D^{\hat{+}b}\phi^{-\hat{-}} = -2i \nabla^{ab}\phi^{-\hat{+}}\,,  \lb{PolRel}
\eea
which can be easily deduced from the previous relations and definitions. Some other useful relations following as Bianchi identities from the basic ones read
\bea
&& [D^{\hat{+}a}, \nabla_{bc}] = \frac{i}{2} \big(\delta^a_b \nabla^-_c  - \delta^a_c \nabla^-_b \big)\phi^{+\hat{+}}\,, \quad
[\nabla^{\hat{-}a}, \nabla_{bc}] = \frac{i}{2} \big(\delta^a_b \nabla^-_c  - \delta^a_c \nabla^-_b \big)\phi^{+\hat{-}}\,, \nn
&& [\nabla^{\pm}_a, \nabla^{bc}] = -\frac{i}{2} \big(\delta^b_a D^{\hat{+}c}  - \delta^c_a D^{\hat{+}b} \big) \phi^{\pm\hat{-}}\,, \lb{DDx} \\
&& [\nabla_{ab}, \nabla^{cd}] = \frac14 \big[\delta^c_a \nabla^-_b \nabla^{\hat{-}d} -  \delta^d_a \nabla^-_b \nabla^{\hat{-}c} - (a\, \leftrightarrow \,b)\big]\phi^{+\hat{+}}\,. \lb{DxDx}
\eea

Now we are armed with all the necessary ingredients for constructing on-shell invariants of $6D, {\cal N}=(1,1)$ SYM theory in ${\cal N}=(1,1)$ harmonic superspace, in
the manifestly gauge invariant and $6D, {\cal N}=(1,1)$ supersymmetric fashion.

\section{On-shell ${\cal N}=(1,1)$ SYM invariants}

\subsection{Generalities}
The building-blocks of all higher-dimension invariants of ${\cal N}=(1,1)$ SYM theory in the bi-harmonic approach are
the superfield strength $\phi^{+\hat{+}}$, its various harmonic projections  obtained by acting on it by the gauge-covariant
harmonic derivatives $\nabla^{\pm\pm}, \nabla^{\hat{\pm}\hat{\pm}}$, as well as the superfields obtained through  acting of the spinor and vector gauge-covariant derivatives
$\nabla^{\pm}_a, D^{\hat{+}a}, \nabla^{\hat{-}a}$ and $\nabla_{ab}$ on the harmonic projections just mentioned. These objects live either on the entire ${\cal N}= (1,1)$ bi-HSS
or on its analytic subspaces.  To ensure gauge invariance, one is led
to take traces of the products of such elementary matrix blocks, each block being valued in the adjoint representation of the gauge group algebra (like $\phi^{+\hat{+}}$).  As shown
in \cite{12}, given some covariant ${\cal N}= (1,1)$ superfield $\Phi (\hat{Z})$,
the transformations of hidden ${\cal N}=(0,1)$ supersymmetry on the ${\cal N}=(1,0)$ superfield constituents of such products, before taking traces, are induced
by the following generic transformation of $\Phi (\hat{Z})$
\bea
\delta_{(0,1)}\Phi (\hat{Z}) = \delta_{stand} \Phi (\hat{Z}) + [\delta_{(1)}^{\rm comp} + \delta_{(2)}^{\rm comp}]\Phi (\hat{Z})\,.\lb{01Gen}
\eea
Here, $\delta_{stand}$ is the standard ${\cal N}=(0,1)$ superfield transformation in the analytic basis induced by the proper shifts of the superspace coordinates,
\bea
\delta_{stand} \Phi (\hat{Z}) = - \left(\epsilon^{\hat{+}}_a \frac{\partial}{\partial \theta^{\hat{+}}_a} + \epsilon^{\hat{-}}_a \frac{\partial}{\partial \theta^{\hat{-}}_a}
+ 2i\epsilon^{\hat{-}}_a \theta^{\hat{+}}_b \partial^{ab} \right)\Phi (\hat{Z}), \lb{stand01}
\eea
while $\delta_{(1)}^{\rm comp}$ and $\delta_{(2)}^{\rm comp}$ are compensating gauge transformations with analytic parameters $\tilde{\Lambda}_{(1)}, \tilde{\Lambda}_{(2)}$
needed to preserve various superfield gauges imposed in the process of solving the constraints \p{ConstrHarm}
\be
\tilde{\Lambda}_{(1)} = 2i\epsilon^{\hat{-}}_a \theta^{\hat{+}}_b {\cal A}^{ab} + {\rm higher \; orders \; in\,} \theta^{\hat +}_b\,, \quad \tilde{\Lambda}_{(2)} =
\epsilon^{-B} q^{+}_B\,, \lb{12Gauge}
\ee
where the higher-order $\theta^{\hat{+}}_a$  terms in $\tilde{\Lambda}_{(1)}$ are of no interest for us here. Actually, the same formula for variation is applicable not only to gauge-covariant expressions
but equally for the various gauge connections. Applying it to the basic entity $\phi^{+\hat{+}}$ defined in \p{explphi}, we find the following ${\cal N}=(0,1)$ transformation
of the basic ${\cal N}=(1,0)$ superfield objects
\bea
\delta_{(0,1)} q^{+ A} = \epsilon^A_b W^{+ b} - [\epsilon^{-B}q^+_B, q^{+ A}]\,, \quad \delta_{(0,1)} W^{+ a} =
-2i\epsilon^A_b\,{\cal D}^{ab} q^{+}_A - [\epsilon^{-B}q^+_B, W^{+ a}]\,. \lb{Hidd}
\eea
These transformations could be as well directly derived from \p{11Susy} by imposing the equations of motion \p{EqMq}, \p{EqMV+q}. For
$q^{-A} = \nabla^{--}q^{+A}$ and $W^{- b} =\nabla^{--}W^{+ b}$ one obtains the same transformation rules.

 An important point is that one can omit all the gauge transformation parts in \p{01Gen}, when constructing the higher-order invariants, since the superfields
 like $\Phi(\hat{Z})$ always appear under the trace. So one can construct actual ${\cal N}=(1,1)$ SYM invariants by integrating $\Phi(\hat{Z})$ over the appropriate invariant
 subspaces of ${\cal N}=(1,1)$ bi-HSS.

 We start by recalling some invariants obtained in \cite{12}. In what follows, the dimension ${\bf d}$ of diverse invariants is understood as the canonical
 dimension (in the mass units) of their component
 Lagrangian. So the dimension of the ``microscopic'' action $S_{(1,1)}$ is ${\bf d=4}$. Note that it makes no sense to ask whether this simplest invariant
 can be represented as an integral over
 ${\cal N}=(1,1)$ superspace since an essential feature of our formalism is that it is on-shell. The relevant ${\cal N}= (1,0)$ superfields
 are assumed to satisfy
 the equations of motion \p{EqMq}, \p{EqMV+q} following just from  $S_{(1,1)}$, so the latter vanishes on the shell of these equations.
 It will be useful to list the dimensions of the basic building blocks:
 \be
 [\nabla^{\pm}_a] = [\nabla^{\hat{-}a}] = [D^{\hat{+}a}] = 1/2\,, \quad [\nabla^{ab}] = 1\,, \quad [\nabla^{\pm\pm}] = [\nabla^{\hat{\pm}\hat{\pm}}] = 0\,, \quad
 [\phi^{+\hat{+}}] = 1\,. \lb{DiM}
 \ee
Also, we list the dimensions of various ${\cal N}=(1,0)$ and ${\cal N}=(1,1)$ superspace integration measures defined in \p{10Integ} - \p{11Int-b}:
\bea
&& [dZ] = -2\,, \quad  [d\zeta^{(-4)}] = -4\,, \lb{10Meas} \\
&& [d\hat{Z}] = 2\,, \quad [d\hat{\zeta}^{(-4,\hat{-}4)}] = -2\,, \quad  [d\hat{\zeta}^{(0,\hat{-}4)}_I] = [d\hat{\zeta}^{(-4, 0)}_{II}] =0\,.\lb{11Meas}
\eea

It is worth to have in mind that, due to the presence of trace in the possible invariants, it is legal to integrate by parts with respect
to the harmonic, spinor and $x$-derivatives, e.g., ${\rm Tr} \nabla^{++} \Omega = D^{++} {\rm Tr} \Omega$, etc. The same is true for the invariants the Lagrangian densities of which are
representable as products of few traces. For what follows, it will be useful to give the general definition of various
invariants as integrals over the full bi-harmonic superspace and its various subspaces
\bea
&& S := \int d\hat{Z}\, L\,, \quad  {\cal S} := \int d\hat{\zeta}^{(-4, \hat{-}4)} {\cal L}^{(+4, \hat{+}4)}\,, \nn
&& {\cal S}_I := \int d\hat{\zeta}_I^{(0, \hat{-}4)} {\cal L}_I^{(0, \hat{+}4)}\,, \quad
{\cal S}_{II} := \int d\hat{\zeta}_{II}^{(-4, 0)} {\cal L}_{II}^{(+4, 0)}\,. \lb{GenInvDef}
\eea

\subsection{Examples}
As promised, we start with recalling the results of \cite{12}.\\

\noindent\underline{\bf d=4}. For completeness, we will start from the case with the canonical dimension of the microscopic ${\cal N}=(1,1)$ SYM action. In this case
the dimension of possible extra invariant is ${\bf d=-2}$, so we obtain
\be
[L] = -4\,, \quad [{\cal L}^{(+4, \hat{+}4)}] = 0\,, \quad [{\cal L}_I^{(0, \hat{+}4)}] =[{\cal L}_{II}^{(+4, 0)}] = -2\,.
\ee
It is impossible to construct such Lagrangian densities from the basic building blocks mentioned in the beginning of this Section,
because all of them have the strictly positive dimensions. \\

\noindent\underline{\bf d=6}. In this case the possible invariants should be dimensionless, so we have
\be
[L] = -2\,, \quad [{\cal L}^{(+4, \hat{+}4)}] = 2\,, \quad [{\cal L}_I^{(0, \hat{+}4)}] =[{\cal L}_{II}^{(+4, 0)}] = 0\,.
\ee
All invariants except the second one are ruled out by the dimensionality reasoning, like in the ${\bf d = 4}$ case. It is easy to be convinced that no
analytic bi-harmonic densities of the dimension 2 and the charges $(+4, \hat{+}4)$ can be constructed from the basic ``bricks''.
So there exist no on-shell ${\cal N} =(1,1)$ invariants with the component Lagrangian density of the canonical dimension $6$. This amounts to the one-loop  finiteness of
$6D\,, \,{\cal N}=(1,1)$ SYM theory. \\

\noindent\underline{\bf d=8}. The relevant invariants should have the dimension ${\bf d=2}$, that implies
\be
[L] = 0\,, \quad [{\cal L}^{(+4, \hat{+}4)}] = 4\,, \quad [{\cal L}_I^{(0, \hat{+}4)}] =[{\cal L}_{II}^{(+4, 0)}] = 2\,.
\ee
Like in the previous case, one can rule out the last two options since no hat-analytic or analytic densities of the dimension 2 and the charges $+4$ or $\hat{+}4$ can be
constructed out of the basic blocks. The first option is dismissed on the dimensionality grounds, so we are left with the second option. The corresponding superfield
single-trace Lagrangian is specified up to an overall constant, so that
\be
{\cal S}_{(8)}^{(1)} \sim {\rm Tr}\,\int d\hat{\zeta}^{(-4, \hat{-}4)}\,(\phi^{+\hat{+}})^4\,.\lb{dim8}
\ee
Note that the Lagrangian in \p{dim8} is analytic only on shell since the first condition in \p{twoAnal} is satisfied with taking into account
the ${\cal N}=(1,0)$ superfield equations of motion. This means that both manifest and hidden supersymmetries of \p{dim8} are fulfilled on shell.
The same property is shared by the double-trace invariant
\be
{\cal S}_{(8)}^{(2)} \sim \int d\hat{\zeta}^{(-4, \hat{-}4)}\,{\rm Tr}(\phi^{+\hat{+}})^2{\rm Tr}(\phi^{+\hat{+}})^2\,.\lb{2dim8}
\ee
Since the $6D, {\cal N}=(1,0)$ HSS supergraph techniques \cite{Q6D1,Q6D2,Q6D3} should yield expressions with {\it off-shell} ${\cal N}=(1,0)$ supersymmetry, the absence
of such invariants indicates the two-loop finiteness of $6D\,, \,{\cal N}=(1,1)$ SYM theory \footnote{The explicit quantum harmonic superfield calculations confirming
this statement were performed in \cite{2La,2Lb}.}.\\

\noindent\underline{\bf d=10}. The dimension of the relevant invariant is ${\bf d=4}$, so the dimensions of the superfield Lagrangian densities are
\be
[L] = 2\,, \quad [{\cal L}^{(+4, \hat{+}4)}] = 6\,, \quad [{\cal L}_I^{(0, \hat{+}4)}] =[{\cal L}_{II}^{(+4, 0)}] = 4\,.
\ee
A simple analysis shows that no proper double-analytic expression of the dimension 6 with the charges $(+4, \hat{+}4)$ exists, even if the equations of motion are assumed.
At the same time, the first and last invariants can be easily constructed:
\bea
S_{(10)} = {\rm Tr}\int d\hat{Z}\, \phi^{+\hat{+}}\phi^{-\hat{-}} \lb{Inv10}
\eea
(another admissible invariant $\sim \phi^{+\hat{-}}\phi^{-\hat{+}}$ is reduced to \p{Inv10} after integration by parts). A single-trace hat-analytic invariant is given by the expression
\bea
{\cal S}_{I(10)} := {\rm Tr}\int d\hat{\zeta}_I^{(0, \hat{-}4)}\,(\phi^{+\hat{+}})^2(\phi^{-\hat{+}})^2 \,. \lb{Inv10Anal1}
\eea

It is interesting that both these invariants can be reduced to each other. Representing
\bea
d\hat{Z} = d\hat{\zeta}_I^{(0, \hat{-}4)} (D^{\hat{+}})^4
\eea
and taking into account that the factor $(D^{\hat{+}})^4$ in \p{Inv10} acts only on $\phi^{-\hat{-}}$ in view of the hat-analyticity of $\phi^{+\hat{+}}$, it is rather straightforward to find,
using, in particular,  the relations \p{PolRel}, that
\bea
(D^{\hat{+}})^4\phi^{-\hat{-}} = -[ [\phi^{+\hat{+}}, \phi^{-\hat{+}}], \phi^{-\hat{+}}]\,.
\eea
Hence,
\bea
S_{(10)} = {\rm Tr}\,\int d\hat{\zeta}_I^{(0,\hat{-}4)}[\phi^{+\hat{+}}, \phi^{-\hat{+}}][\phi^{+\hat{+}}, \phi^{-\hat{+}}]\,.
\eea
On the other hand, using the integration by parts with respect to $\nabla^{\hat{+}\hat{+}}$, one can show that
\bea
S_{(10)} = -6{\rm Tr}\,\int d\hat{\zeta}_I^{(0,\hat{-}4)} (\phi^{+\hat{+}})^2(\phi^{-\hat{+}})^2\,. \lb{04Repr}
\eea
So,
\bea
S_{(10)} = -6 {\cal S}_{I(10)}\,.
\eea

Quite analogously, representing $d\hat{Z} = d\hat{\zeta}_{II}^{(-4,0)} (D^+)^4$ and taking advantage of the property that $D^+_a \rightarrow \nabla^+_a$ under trace, one can show that
\bea
S_{(10)} = {\rm Tr}  \int d\hat{\zeta}_{II}^{(-4,0)} [\phi^{+\hat{+}}, \phi^{+\hat{-}}] [\phi^{+\hat{+}}, \phi^{+\hat{-}}] =
-6{\rm Tr}  \int d\hat{\zeta}_{II}^{(-4,0)}(\phi^{+\hat{+}})^2(\phi^{+\hat{-}})^2\,.\lb{40Repr}
\eea

Note an essential difference between the representations \p{04Repr} and \p{40Repr}. When passing from \p{Inv10} to \p{04Repr}, no use of the
${\cal N}=(1,0)$ superfield equations was made, so ${\cal N}=(1,0)$ supersymmetry can be still treated as an off-shell supersymmetry, the equations of motion
are needed only for demonstrating the hidden ${\cal N}=(0,1)$ supersymmetry. The Lagrangian in \p{04Repr} is manifestly hat-analytic since the superfield strengths
$\phi^{\pm \hat{+}}$ do not depend on $\theta^{\hat{-}}_a$. On the other hand, when
passing to the representation \p{40Repr}, it is imperative to use the equations of motion, because, e.g., the property $\nabla^+_a\phi^{+\hat{+}} = 0$ is valid
only on shell. So the Lagrangian in \p{40Repr} is ${\cal N}=(1,0)$ analytic only on shell, and ${\cal N}=(1,0)$ supersymmetry is valid also on shell,
like in the case of ${\bf d=8}$ invariant \p{dim8}. In other words, the equivalency between the initial full bi-harmonic superspace single-trace invariant \p{Inv10} and the invariant \p{04Repr}
is valid only on shell.  Thus \p{04Repr} provides another example of ${\cal N}=(1,1)$ invariants in which both ${\cal N}=(1,0)$ and ${\cal N}=(0,1)$ supersymmetries
are on shell. This peculiarity  was not noticed in \cite{12}.

As for the double-trace invariants of the dimension ${\bf d=10}$,  they can be defined only in the $3/4$ analytic superspaces,
\bea
&& {\cal S}_{(10)I}^{(2)} \sim \int d\hat{\zeta}_I^{(0,\hat{-}4)} {\rm Tr}\,(\phi^{+\hat{+}} \phi^{-\hat{+}}){\rm Tr}\,(\phi^{+\hat{+}} \phi^{-\hat{+}})\,, \lb{Double10dI} \\
&& {\cal S}_{(10)II}^{(2)} \sim \int d\hat{\zeta}_{II}^{(-4, 0)} {\rm Tr}\,(\phi^{+\hat{+}} \phi^{+\hat{-}}){\rm Tr}\,(\phi^{+\hat{+}} \phi^{+\hat{-}})\,. \lb{Double10dII}
\eea
Like in the single-trace case, the Lagrangian density in \p{Double10dII} is analytic only on shell, and the same is true for the realization of ${\cal N}=(1,0)$
supersymmetry in it. So \p{Double10dII} cannot appear as a three-loop divergence in the explicitly ${\cal N} = (1,0)$ off-shell supersymmetric perturbation calculations.
However, this noticeable property does not fully explain the absence of the double-trace counterterms in the amplitude calculations for ${\cal N}=(1,1)$ SYM
\cite{13,14,Bern:2012uf,BGRV,Bjornsson:2010wm,Bjornsson:2010wu}, since the alternative
double-trace invariant \p{Double10dI} is off-shell ${\cal N}=(1,0)$ supersymmetric and some extra reasoning is yet needed to rule out it \cite{Bossard:2009mn}.

\section{Invariants of dimension ${\bf d=12}$}
In this case the relevant component actions should have the canonical dimension ${\bf d=6}$. Respectively, for dimensions
of various superfield Lagrangians defined in \p{GenInvDef} we obtain
\be
[L] = 4\,, \quad [{\cal L}^{(+4, \hat{+}4)}] = 8\,, \quad [{\cal L}_I^{(0, \hat{+}4)}] =[{\cal L}_{II}^{(+4, 0)}] = 6\,.\lb{Dimd12}
\ee

First of all,  it is important to realize that it is impossible to construct, out of the elementary ``bricks''  $\phi^{\pm\hat{\pm}}, \phi^{\mp\hat{\pm}}$
and those obtained from them through action of the covariant differential operators $\nabla^{\pm}_a, D^{\hat{+}a}, \nabla^{\hat{-}a}$ and $\nabla_{ab}$,
the gauge invariant and manifestly analytic objects possessing the charges $(+4, \hat{+}4)$, $(+4, 0)$ or  $(0, \hat{+}4)$ and canonical dimensions indicated in \p{Dimd12}\footnote{In principle,
some Chern-Simons type not manifestly gauge-invariant densities with these charge and dimension assignments could be imagined
(see, e.g. \cite{CS1,CS2,CS3});
however, here we do not consider such possibilities.}. So we are left with the chargeless general bi-HSS superfield densities $L$ of the dimension $4$  as the only
candidates for the invariants we are interested in.

\subsection{Invariants without derivatives}
We start with the invariants containing only superfield strengths $\phi^{\pm\hat{\pm}}, \phi^{\mp\hat{\pm}}$. One can construct ten chargeless superfield invariants of the dimension 4
from these quantities
\bea
&& J_1 = {\rm Tr}\, \phi^{+\hat{+}} \phi^{+\hat{+}}\phi^{-\hat{-}}\phi^{-\hat{-}}\,, \quad   J_2 = {\rm Tr}\, \phi^{+\hat{+}} \phi^{-\hat{-}}\phi^{+\hat{+}}\phi^{-\hat{-}}\,, \nn
&& J_3 = {\rm Tr}\, \phi^{+\hat{-}} \phi^{+\hat{-}}\phi^{-\hat{+}}\phi^{-\hat{+}}\,, \quad J_4 = {\rm Tr}\, \phi^{+\hat{-}} \phi^{-\hat{+}}\phi^{+\hat{-}}\phi^{-\hat{+}}\,, \lb{Jays} \\
&&I_1 = {\rm Tr}\, \phi^{+\hat{+}} \phi^{+\hat{-}}\phi^{-\hat{+}}\phi^{-\hat{-}}\,, \quad   I_2 = {\rm Tr}\, \phi^{+\hat{+}} \phi^{-\hat{+}}\phi^{-\hat{-}}\phi^{+\hat{-}}\,, \nn
&& I_3 = {\rm Tr}\, \phi^{+\hat{+}} \phi^{-\hat{-}}\phi^{+\hat{-}}\phi^{-\hat{+}}\,, \quad I_4 = {\rm Tr}\, \phi^{+\hat{+}} \phi^{-\hat{-}}\phi^{-\hat{+}}\phi^{+\hat{-}}\,, \nn
&& I_5 = {\rm Tr}\, \phi^{+\hat{+}} \phi^{+\hat{-}}\phi^{-\hat{-}}\phi^{-\hat{+}}\,, \quad I_6 = {\rm Tr}\, \phi^{+\hat{+}} \phi^{-\hat{+}}\phi^{+\hat{-}}\phi^{-\hat{-}}\,. \lb{Ays}
\eea

Some of them are related via the integration by parts with respect to harmonic derivatives. To extract the irreducible set, one needs to take off some derivatives with ``hat''
and/or  without ``hats''
from some superfield factors in \p{Jays} and \p{Ays} and place them on the other factors in the same expression, using the relations \p{DefProjphi}, \p{Relphi}
and the on-shell harmonic constraints
\p{HarmConstrphi}, \p{Analphi2}. This can be done in a few equivalent ways. It will be convenient for us to consecutively take off the derivatives
$\nabla^{--}, \nabla^{\hat{-}\hat{-}}$ from all
negatively charged factors in the densities $J_1, \ldots, J_4$. For instance, representing the third factor $\phi^{-\hat{-}}$ in $J_1$ under trace as $ \phi^{-\hat{-}} =
\nabla^{--}\phi^{+\hat{-}} = \nabla^{\hat{-}\hat{-}}\phi^{-\hat{+}}$ and integrating by parts with respect to $\nabla^{--}$ and $\nabla^{\hat{-}\hat{-}}$, we obtain, respectively,
\bea
&& J_1 \simeq - {\rm Tr}\, \phi^{-\hat{+}} \phi^{+\hat{+}}\phi^{+\hat{-}}\phi^{-\hat{-}} - {\rm Tr}\, \phi^{+\hat{+}} \phi^{-\hat{+}}\phi^{+\hat{-}}\phi^{-\hat{-}} = - I_5 - I_6\,, \nn
&& J_1 \simeq - {\rm Tr}\, \phi^{+\hat{-}} \phi^{+\hat{+}}\phi^{-\hat{+}}\phi^{-\hat{-}} - {\rm Tr}\, \phi^{+\hat{+}} \phi^{+\hat{-}}\phi^{-\hat{+}}\phi^{-\hat{-}} = - I_2 - I_1\,, \lb{Exs}
\eea
etc, where $\simeq$ means ``up to a total harmonic derivative''. In this way we obtain the following set of linear equations (we replace $\simeq$ by $=$ for brevity)
\bea
&& J_1 +I_5 + I_6=0\,, \; J_1 +I_2 + I_1=0\,, \; J_1 +I_3 + I_2=0\,, \; J_1 +I_4 + I_5=0\,, \lb{J1eq} \\
&&J_2 +I_4 + I_1=0\,, \quad J_2 +I_3 + I_6=0\,,\lb{J2eq} \\
&& J_3 +I_2 + I_6=0\,, \; J_3 +I_3 + I_5=0\,, \; J_3 +I_1 + I_5=0\,, \; J_3 +I_2 + I_4=0\,,\lb{J3eq} \\
&&J_4 +I_1 + I_4=0\,, \quad J_4 +I_3 + I_6=0\,.\lb{J2eq2}
\eea
It is direct to check that applying the same procedure to the densities $I_1, \ldots,  I_6$ produces no new relations. Moreover, the system \p{J1eq} - \p{J2eq2} is underdetermined and
in fact amounts to eight independent equations for ten unknowns. As a result, we are left with the two independent densities. As such it is convenient to choose $J_1$ and $J_2$:
\bea
I_1 = I_3 = I_4 = I_6 = -\frac12 J_2 \,, \quad I_2 = I_5 = \frac12 J_2 - J_1, \quad J_3 = J_1\,,\; J_4 = J_2\,, \lb{IndepJ}
\eea
where it is assumed as before that the equalities are valid up to total $\nabla^{--}$ or $\nabla^{\hat{-}\hat{-}}$ derivatives.

So in dimension ${\bf d=12}$ we have two independent single-trace superfield invariants without derivatives,
\bea
S_{(12)I}^{(1)} \sim {\rm Tr} \int d\hat{Z}\, \phi^{+\hat{+}} \phi^{+\hat{+}}\phi^{-\hat{-}} \phi^{-\hat{-}}\,, \quad  S_{(12)II}^{(1)} \sim {\rm Tr} \int d\hat{Z}\,
\phi^{+\hat{+}} \phi^{-\hat{-}}\phi^{+\hat{+}} \phi^{-\hat{-}}\,. \lb{d12Inv}
\eea
Analogously, one can define two independent double-trace invariants of the same type:
\bea
&& S_{(12)I}^{(2)} \sim \int d\hat{Z}\, {\rm Tr}\,(\phi^{+\hat{+}} \phi^{+\hat{+}})\,{\rm Tr}\,(\phi^{-\hat{-}} \phi^{-\hat{-}})\,, \nn
&& S_{(12)II}^{(2)} \sim \int d\hat{Z}\,
{\rm Tr} (\phi^{+\hat{+}} \phi^{-\hat{-}})\,{\rm Tr}\,(\phi^{+\hat{+}} \phi^{-\hat{-}})\,. \lb{d12InvD}
\eea

\subsection{Invariants with derivatives}
Next we consider possible superfield densities containing spinor and/or vector covariant derivatives. The only possibilities to ensure the total dimension four are
the densities with three covariant strengths and two spinor derivatives, or densities with two
strengths and either two vector, or four spinor, or two spinor and one vector derivatives properly distributed among these strengths. It is clear that the double-trace
invariants of this kind do not exist in view of the tracelessness of the gauge algebra generators.

We start with the three-strength case. Taking into account the on-shell analyticities of various harmonic projections of the superfield strength and the possibility to integrate
by parts under the trace, we can construct two options
\bea
S_{(12)(3)} = {\rm Tr}\,\int d \hat{Z}\,D^{\hat{+}a} \phi^{-\hat{-}}\nabla^+_a\phi^{-\hat{-}}\phi^{+\hat{+}},  \,
S_{(12)(3)}' = {\rm Tr}\,\int d \hat{Z}\,\nabla^{+}_a \phi^{-\hat{-}} D^{\hat{+}a}\phi^{-\hat{-}}\phi^{+\hat{+}}. \lb{Set1}
\eea
Integrating by parts with respect to $D^{\hat{+}a}$ in the first expression and using $\{\nabla^+_a, D^{\hat{+}a} \} = 4\phi^{+\hat{+}}$, we obtain
\bea
&&{\rm Tr}\,D^{\hat{+}a} \phi^{-\hat{-}}\nabla^+_a\phi^{-\hat{-}}\phi^{+\hat{+}} \simeq -{\rm Tr}\,\phi^{-\hat{-}}D^{\hat{+}a}\nabla^+_a\phi^{-\hat{-}}\phi^{+\hat{+}} \nn
&& = {\rm Tr}\,\phi^{-\hat{-}}\nabla^+_aD^{\hat{+}a}\phi^{-\hat{-}}\phi^{+\hat{+}} -4{\rm Tr}\,\phi^{-\hat{-}}[\phi^{+\hat{+}}, \phi^{-\hat{-}}]\phi^{+\hat{+}} \nn
&& \simeq -{\rm Tr}\,\nabla^{+}_a \phi^{-\hat{-}} D^{\hat{+}a}\phi^{-\hat{-}}\phi^{+\hat{+}} + 4{\rm Tr}\,\big(\phi^{+\hat{+}}\phi^{+\hat{+}} \phi^{-\hat{-}}\phi^{-\hat{-}} -
\phi^{+\hat{+}}\phi^{-\hat{-}} \phi^{+\hat{+}}\phi^{-\hat{-}}\big), \lb{Equiv1}
\eea
whence we conclude that $S_{(12)(3)} = -S_{(12)(3)}'$ up to the difference of the previous invariants $\sim \big[S_{(12)I}^{(1)} - S_{(12)II}^{(1)}]$. So we are left with
only one independent invariant of the type \p{Set1}.

For constructing densities with two superfield strengths there are much more opportunities, but, once again, they are reduced, modulo the
invariants already presented, to the following single invariant
\bea
S_{(12)(2)} \sim {\rm Tr}\,\int d\hat{Z} \nabla_{ab}\phi^{-\hat{-}}\nabla^{ab} \phi^{+\hat{+}}\,. \lb{Twostrength}
\eea

As an example of such a reduction, consider, e.g., the density
\be
{\rm Tr}\,\nabla^+_a\phi^{-\hat{-}} \nabla^-_b \nabla^{ab} \phi^{+\hat{+}}\,. \lb{Two1}
\ee
An evident chain of transformations involving integrations by parts, yields
\bea
&& {\rm Tr}\,\nabla^+_a\phi^{-\hat{-}} \nabla^-_b \nabla^{ab} \phi^{+\hat{+}} \simeq -{\rm Tr}\,\phi^{-\hat{-}} \nabla^+_a\nabla^-_b \nabla^{ab} \phi^{+\hat{+}} =
{\rm Tr}\,\phi^{-\hat{-}} \nabla^-_b \nabla^+_a \nabla^{ab} \phi^{+\hat{+}}\nn
&&-\, 2i\,{\rm Tr}\,\phi^{-\hat{-}}\nabla_{ab}\nabla^{ab}\phi^{+\hat{+}} \simeq 2i\, {\rm Tr}\,\nabla_{ab}\phi^{-\hat{-}}\nabla^{ab} \phi^{+\hat{+}}\,,
\eea
where we made use of the relation $\nabla^-_b\phi^{-\hat{-}} = 0\,$. In a similar way one can handle the density
\be
{\rm Tr}\,D^{\hat{+}a}\phi^{-\hat{-}} \nabla^{\hat{-}b}\nabla_{ab} \phi^{+\hat{+}}\,.
\ee

The density
\be
{\rm Tr}\,\nabla^+_a\phi^{-\hat{-}} \nabla^{ab} \nabla^-_b \phi^{+\hat{+}} \lb{Two2}
\ee
can be reduced to \p{Two1} by using the commutation relations \p{DDx},
\bea
&&{\rm Tr}\,\nabla^+_a\phi^{-\hat{-}} \nabla^{ab} \nabla^-_b \phi^{+\hat{+}} =
{\rm Tr}\,\nabla^+_a\phi^{-\hat{-}} \nabla^-_b \nabla^{ab} \phi^{+\hat{+}} - \frac{3i}2{\rm Tr}\,\nabla^+_a\phi^{-\hat{-}}[D^{\hat{+}a}\phi^{-\hat{-}}, \phi^{+\hat{+}}] \nn
&&  \simeq {\rm Tr}\,\nabla^+_a\phi^{-\hat{-}} \nabla^-_b \nabla^{ab} \phi^{+\hat{+}}  -6i \big(\phi^{+\hat{+}}\phi^{+\hat{+}}\phi^{-\hat{-}}\phi^{-\hat{-}} -
\phi^{+\hat{+}}\phi^{-\hat{-}}\phi^{+\hat{+}}\phi^{-\hat{-}}\big)  \,, \lb{Two2a}
\eea
where we made use of the relation \p{Equiv1}.

Two more examples are the following `trial' densities
\bea
U_I = \varepsilon^{abcd} {\rm Tr}\,\nabla^+_a\nabla^+_b\phi^{-\hat{+}}\nabla^-_c\nabla^-_d\phi^{+\hat{-}}\,, \quad U_{II} =
\nabla^+_a\nabla^+_b\phi^{-\hat{-}} D^{\hat{+}a}D^{\hat{+}b}\phi^{-\hat{-}}\,.
\eea
Using the relations \p{PolRelb}, it is easy to reduce them to the expressions
\be
U_I = -8{\rm Tr}\,\nabla_{ab}\phi^{+\hat{+}}\nabla^{ab}\phi^{-\hat{-}}\,, \quad U_{II} = -4 \nabla_{ab}\phi^{+\hat{-}}\nabla^{ab}\phi^{-\hat{+}}\,,
\ee
which are related to each other via integration by parts with respect to the harmonic derivative $\nabla^{++}$. All other possible expressions, e.g.,
$\sim \varepsilon_{abcd} {\rm Tr}\,D^{\hat{+}a} D^{\hat{+}b} \phi^{-\hat{-}}\nabla^{\hat{-}c}\nabla^{\hat{-}d}\phi^{+\hat{+}}\,,$ can be transformed in
a similar manner.\\

For reader's convenience, we finish this section by listing the complete set of independent ${\cal N}=(1,1)$ superfield invariants corresponding to the component Lagrangian densities
of the canonical dimension ${\bf d = 12}$ .\\

\noindent{\bf Single-trace invariants}
\bea
&& S_{(12)I}^{(1)} \sim {\rm Tr} \int d\hat{Z}\, \phi^{+\hat{+}} \phi^{+\hat{+}}\phi^{-\hat{-}} \phi^{-\hat{-}}\,, \quad  S_{(12)II}^{(1)} \sim {\rm Tr} \int d\hat{Z}\,
\phi^{+\hat{+}} \phi^{-\hat{-}}\phi^{+\hat{+}} \phi^{-\hat{-}}\,, \nn
&& S_{(12)(3)}^{(1)} \sim {\rm Tr}\int d \hat{Z}\,D^{\hat{+}a} \phi^{-\hat{-}}\nabla^+_a\phi^{-\hat{-}}\phi^{+\hat{+}}, \nn
&& S_{(12)(2)}^{(1)} \sim {\rm Tr}\int d\hat{Z}\, \nabla_{ab}\phi^{-\hat{-}}\nabla^{ab} \phi^{+\hat{+}}\,. \lb{AlloneTrace}
\eea

\noindent{\bf Double-trace invariants}
\bea
&& S_{(12)I}^{(2)} \sim \int d\hat{Z}\, {\rm Tr}\,(\phi^{+\hat{+}} \phi^{+\hat{+}})\,{\rm Tr}\,(\phi^{-\hat{-}} \phi^{-\hat{-}})\,, \nn
&& S_{(12)II}^{(2)} \sim \int d\hat{Z}\,
{\rm Tr} (\phi^{+\hat{+}} \phi^{-\hat{-}})\,{\rm Tr}\,(\phi^{+\hat{+}} \phi^{-\hat{-}})\,. \lb{All2Trace}
\eea

It is worth to point out that, as distinct from the ${\bf d=10}$ single-trace invariant, the ${\bf d=12}$ ones cannot be equivalently rewritten as integrals of the manifestly
analytic Lagrangian densities over the appropriate analytic subspaces. It would be interesting to understand possible implications of this property for the structure
of quantum corrections in ${\cal N}=(1,1)$ SYM theory. Anyway, in all these invariants ${\cal N}=(1,0)$ supersymmetry can be treated as an off-shell one and all
of them could in principle appear as divergences in the four-loop order of $6D, {\cal N}=(1,0)$ HSS superfield perturbation calculations.
Note that in the abelian case all invariants without derivatives become identical to each other.

\section{Passing to ${\cal N}=(1,0)$ superfields}
Here we consider the ${\cal N}=(1,0)$ superfield form of some invariants listed in the previous sections. For simplicity, we will confine our attention to the contributions
of the ${\cal N}=(1,0)$ gauge superfield and  ignore the hypermultiplet pieces. So our input for the basic ${\cal N}=(1,1)$ superfields defined in \p{explphi}, \p{HatV++}
will be the following
\bea
\phi^{+\hat{+}} \; \rightarrow \; -\theta^{\hat{+}}_a W^{+ a}\,, \quad V^{\hat{+}\hat{+}} \;\rightarrow \; i\theta^{\hat{+}}_a\theta^{\hat{+}}_b\,{\cal A}^{ab}\,. \lb{Ansatz}
\eea
We also obtain
\be
\phi^{-\hat{+}} = \nabla^{--}\phi^{+\hat{+}} \; \rightarrow \; -\theta^{\hat{+}}_a W^{- a}\,.
\ee

To compute the remaining ${\cal N}=(1,1)$ SYM quantities, we need the explicit expression for $V^{\hat{-}\hat{-}}$ which is defined by the hat-flatness
condition \p{2ndflatness}. It is easy to see from the latter that the $\theta^{\hat{-}}_a$ expansion of $V^{\hat{-}\hat{-}}$ starts with the second-order terms. So
we parametrize $V^{\hat{-}\hat{-}}$ as
\bea
V^{\hat{-}\hat{-}} = i\theta^{\hat{-}}_a \theta^{\hat{-}}_b\,v^{ab} + \Psi^{\hat{-}3 d}\,v^{\hat{+}}_d + \Psi^{\hat{-}4}\,v^{\hat{+}2}\,, \lb{ExpVhatmin}
\eea
where the coefficients are functions of $(\theta^{\hat{+}}_a, u^{\hat{\pm} A})$ and  of the  ${\cal N}=(1,0)$ coordinates $Z = (x^{ab}, \theta^{\pm a}, u^{\pm i})$.
From \p{2ndflatness} we obtain the following equations for these coefficients
\bea
&& \theta^{\hat{+}}_a \big({\cal A}^{ab} - v^{ab} \big) = 0\,, \quad \nabla^{\hat{+}\hat{+}} v^{ab} - 3i \varepsilon^{abcd} \theta^{\hat{+}}_c v^{\hat{+}}_d
- i  \theta^{\hat{+}}_c \theta^{\hat{+}}_d\partial^{ab}{\cal A}^{cd} = 0\,, \nn
&& \nabla^{\hat{+}\hat{+}}v^{\hat{+}}_d + 4\,\theta^{\hat{+}}_d\,v^{\hat{+}2} = 0\,, \qquad \nabla^{\hat{+}\hat{+}}v^{\hat{+}2} = 0\,. \lb{EqsVhatmin}
\eea
These equations fully fix the $(\theta^{\hat{+}}_a, u^{\hat{\pm} A})$ dependence of the coefficients and we obtain the rather simple expression
for $V^{\hat{-}\hat{-}}$ in the approximation considered:
\bea
V^{\hat{-}\hat{-}} = i\theta^{\hat{-}}_a \theta^{\hat{-}}_b\,{\cal A}^{ab} + \frac13\,\Psi^{\hat{-}3 d}\theta^{\hat{+}}_a [{\cal D}^{ab}, {\cal D}_{db}] +
\frac{i}{12}\Psi^{\hat{-}4}\,\theta^{\hat{+}}_a\theta^{\hat{+}}_b\,[{\cal D}^{ac}, [{\cal D}^{bd}, {\cal D}_{cd}]]\,, \lb{SolVhatmin}
\eea
where ${\cal D}_{ab} = \partial_{ab} + {\cal A}_{ab} = \frac12 \varepsilon_{abcd}{\cal D}^{cd}$ and
\be
[{\cal D}_{ab}, {\cal D}_{cd}] = {\cal D}_{ab}{\cal A}_{cd} -\partial_{cd}{\cal A}_{ab}\,. \lb{CovStr}
\ee

In what follows we will be interested in the on-shell form of the invariants. The coefficients in \p{SolVhatmin} are simplified on shell and can be
expressed through the spinor superfield strengths $W^{\pm a}$. We can use one of the equivalent forms of the equations of motion for $W^{\pm a}$,
\be
D^+_a W^{- b} = - {\cal D}^-_a W^{+ b}\,,
\ee
and represent, using the relation \p{nabla-nabla2}, the second coefficient in the expansion \p{SolVhatmin}  as
\be
[{\cal D}^{ab}, {\cal D}_{db}] = \frac 12 D^{+}_d W^{-a} = -\frac12 {\cal D}^{-}_d W^{+a}\,. \lb{Second}
\ee
Also, after simple algebra, using the on-shell cyclic identity \p{cyclic}, we derive
\bea
[{\cal D}^{[ac}, [{\cal D}^{b]d}, {\cal D}_{cd}]] = \frac{i}2 \{ W^{+[a}, W^{-b]} \}\,. \lb{Second2}
\eea

Now one can compute the total vector connection
\bea
{\cal V}^{ab} \rightarrow  {\cal A}^{ab} -\frac{i}{2}\varepsilon^{abcd} \theta^{\hat{-}}_c \theta^{\hat{+}}_g\,D^{+}_d W^{-g}
+\frac{i}{4}\varepsilon^{abcd}\theta^{\hat{-}}_c \theta^{\hat{-}}_d\theta^{\hat{+}}_g\theta^{\hat{+}}_f \{ W^{+g}, W^{-f} \}, \lb{VabEx}
\eea
and also the rest of the harmonic projections of the covariant superfield strength
\bea
\phi^{+\hat{-}} &\rightarrow& -\theta^{\hat{-}}_a W^{+ a} - i\theta^{\hat{-}}_b\theta^{\hat{-}}_c\theta^{\hat{+}}_a {\cal D}^{bc} W^{+ a}
- \frac16 \Psi^{\hat{-}3 d}\theta^{\hat{+}}_a\theta^{\hat{+}}_b D^{{+}}_d \{W^{-a}, W^{+ b}\} \nn
&&+\, \frac{1}{24\cdot 6}\Psi^{\hat{-}4}\Psi^{\hat{+}3 d}\varepsilon_{abcd}
[\{W^{+a}, W^{- b}\}, W^{+ c}]\,,  \lb{fi+-Ex}\\
\phi^{-\hat{-}} &\rightarrow& -\theta^{\hat{-}}_a W^{- a} - i\theta^{\hat{-}}_b\theta^{\hat{-}}_c \theta^{\hat{+}}_a{\cal D}^{bc} W^{- a}
- \frac16 \Psi^{\hat{-}3 d}\theta^{\hat{+}}_a\theta^{\hat{+}}_b [D^{{+}}_d W^{-a}, W^{- b}] \nn
&&+\, \frac{1}{24\cdot 6}\Psi^{\hat{-}4}\Psi^{\hat{+}3 d}\varepsilon_{abcd}
[\{W^{+a}, W^{- b}\}, W^{- c}]\,. \lb{fi--Ex}
\eea

As the first example of passing to ${\cal N}= (1,0)$ superfield notation, we rewrite the on-shell invariant \p{40Repr}. Using the representation
\be
d\hat{\zeta}_{II}^{(-4,0)} = d{\zeta}^{(-4)} d\hat{u} (D^{\hat{-}})^4 (D^{\hat{+}})^4 \,,
\ee
and the relations $(D^{\hat{\pm}})^4 \Psi^{\hat{\mp} 4} = -4!$ (plus a total $x$-derivative), we find
\bea
S_{(10)} &\rightarrow& \int d{\zeta}^{(-4)}\,{\cal L}^{+4}_{(10)}, \nn
{\cal L}^{+4}_{(10)} &=& \varepsilon_{abcd} {\rm Tr} \big\{ \big(W^{+ a} W^{+ b} W^{+ e} + W^{+ e}W^{+ a} W^{+ b}\big)[ {\cal D}^-_{e} W^{+c}, W^{+ d}] \nn
&& -\, 2 W^{+ a} W^{+ b} {\cal D}_{gf} W^{+ c} {\cal D}^{gf}W^{+ d} \big\}.\lb{10new}
\eea
It is straightforward  to check the implicit on-shell analyticity of ${\cal L}^{+4}_{(10)}$,
\bea
D^{+}_{a}{\cal L}^{+4}_{(10)} = 0\,. \nonumber
\eea
This highlights the property that the ${\bf d=10}$ single-trace invariant in the form \p{40Repr} is both ${\cal N}=(1,0)$ and ${\cal N}=(0,1)$ supersymmetric
only on shell. Note that the same Lagrangian, up to a total derivative, can be rewritten as the trace of two (anti)commutators
\bea
{\cal L}^{+4}_{(10)} &=& \frac13\varepsilon_{abcd} {\rm Tr} \big([D^+_f W^{-a}, W^{+ b}][\{W^{+ c}, W^{+ f}\},  W^{+ d}] \nn
&& +\, \{W^{+ a}, {\cal D}_{gf} W^{+ b}\} \{ W^{+ c}, {\cal D}^{gf}W^{+ d}\} \big),  \lb{10new1}
\eea
from which it follows, in particular,  that this invariant vanishes in the abelian limit.

Analogously, one can find the ${\cal N}=(1,0)$ superfield core of the on-shell double-trace invariant \p{Double10dII}:
\be
{\cal S}_{(10)II}^{(2)} \,\rightarrow \,\int d{\zeta}^{(-4)}{\cal L}^{+4}_{(10)II}\,, \nonumber
\ee
\bea
&& {\cal L}^{+4}_{(10)II} =\varepsilon_{abcd} {\rm Tr} \big(W^{+a} W^{+ b}\big) {\rm Tr} \big({\cal D}_{gf} W^{+c}{\cal D}^{gf} W^{+d} + W^{+ f}[D^+_f W^{-c}, W^{+ d}]\big) \nn
&& \simeq -2\varepsilon_{abcd}{\rm Tr} \big({\cal D}_{gf} W^{+a} W^{+b}\big){\rm Tr} \big({\cal D}^{gf} W^{+c} W^{+ d}\big). \lb{10supII}
\eea
The on-shell analyticity of this Lagrangian can also be easily checked. In the abelian limit this expression is reduced to
\be
\sim \varepsilon_{abcd} \partial_{gf} W^{+a} W^{+b}\partial^{gf} W^{+c} W^{+ d}\,,
\ee
that is a total derivative, when taking account of the condition \p{DirDea}.

Let us proceed to ${\bf d=12}$ invariants. We use $d\hat{Z} = d Z d\hat{u}(D^{\hat{-}})^4(D^{\hat{+}})^4$. For the single-trace invariants without derivatives,
$S^{(1)}_{(12)I}$ and $S^{(1)}_{(12)II}$ defined in \p{d12Inv}, we find
\bea
S^{(1)}_{(12)I,II} &\rightarrow & \int dZ\, L^{(1)}_{(12)I, II}\,, \lb{SI,II} \\
L^{(1)}_{(12)I} &=& -\varepsilon_{abcd}{\rm Tr}\big\{ \big(W^{+a}W^{+b}W^{-f}W^{-c} + W^{-a}W^{-f}W^{+b}W^{+c} \nn
&&+\, W^{-a}W^{+b}W^{+c}W^{-f} + W^{-f}W^{+a}W^{+b}W^{-c} \big){D}^+_f W^{-d} \nn
&&+\,  2 W^{+a}W^{+b} {\cal D}_{fg} W^{-c}{\cal D}^{fg} W^{-d}\big\} \nn
&=& - 2\varepsilon_{abcd}{\rm Tr}\big\{ \big(W^{-a}W^{+b}W^{+c}W^{-f} + W^{-f}W^{+a}W^{+b}W^{-c}\big){D}^+_f W^{-d} \nn
&& +\, W^{+a}W^{+b} {\cal D}_{fg} W^{-c}{\cal D}^{fg} W^{-d}\big\}, \lb{LI} \\
L^{(1)}_{(12)II} &=& 2\varepsilon_{abcd}{\rm Tr}\big\{\big(W^{-a}W^{+b}W^{-f}W^{+c} + W^{+a}W^{-f}W^{+b}W^{-c} \big){D}^+_f W^{-d} \nn
&&-\,  W^{+a}{\cal D}_{fg} W^{-b}W^{+c} {\cal D}^{fg} W^{-d}\big\}. \lb{LII}
\eea
When passing to another representation of $L^{(1)}_{(12)I}$ in \p{LI}, we integrated by parts with respect to $D^+_f$ and used the equations of motion once again.

Quite analogously, one can compute the ${\cal N}=(1,0)$ SYM cores of the double-trace invariants
$S^{(2)}_{(12)I}$ and $S^{(2)}_{(12)II}$. Integrating by parts and using the on-shell condition \p{Dsquare}, we find
\bea
S^{(2)}_{(12)I,II} &\rightarrow & \int dZ\, L^{(2)}_{(12)I, II}\,, \nn
L^{(2)}_{(12)I} &=& 4\,\varepsilon_{abcd}{\rm Tr} \big(W^{+ a}{\cal D}_{gf} W^{+ b}\big){\rm Tr} \big(W^{-c}{\cal D}^{gf} W^{-d}\big), \lb{doubleIa} \\
L^{(2)}_{(12)II} &=& -2 \,\varepsilon_{abcd}\big\{{\rm Tr}\big(W^{+a}{\cal D}_{fg}W^{-b}\big){\rm Tr}\big(W^{+c}{\cal D}^{fg}W^{-d}\big) \nn
&& +\, {\rm Tr}\big(W^{-f}W^{+a}\big) {\rm Tr}\big(\{W^{+b}, W^{-c}\}D^+_f W^{-d}\big)\big\}.   \lb{doubleIIa}
\eea

Recall that in the abelian limit these invariants become identical to each other up to numerical coefficients. They can be reduced to the expression
\bea
S^{\rm (abel)}_{(12)} = \varepsilon_{abcd}\int dZ\, \partial^{gf}W^{+a}\partial_{gf}W^{+ b}W^{-c}W^{-d}\,,\lb{Abel}
\eea
which is not a total derivative even on shell, with $\partial_{gf}\partial^{gf}W^{\pm d} =0$. Hence this sort of invariants could be relevant
to the Coulomb branch of the theory, with  the original gauge symmetry being broken down to some abelian subgroup (e.g., Cartan subgroup).

Before going to the invariants with the spinor and vector derivatives, it will be convenient to define the set of ${\cal N} = (1,0)$ SYM invariants
in terms of which the ${\cal N} = (1,0)$ SYM cores of the single-trace on-shell ${\cal N} = (1,1)$ invariants considered here admit a succinct compact expression.
Using the on-shell conditions and integrating by parts with respect to various derivatives (including the harmonic ones), it is possible to show that
there exist only five independent structures of this kind. It is convenient to choose them as
\bea
&& E_1 = \varepsilon_{abcd} {\rm Tr} \int dZ\, W^{+ a} W^{+ b} W^{-c} W^{-f} D^{+}_f W^{- d}\,, \lb{INV1} \\
&& E_2 = \varepsilon_{abcd} {\rm Tr} \int dZ\, W^{- a} W^{+ b} W^{+c} W^{-f} D^{+}_f W^{- d}\,, \lb{INV2} \\
&& E_3 = \varepsilon_{abcd} {\rm Tr} \int dZ\, W^{- a} W^{-f} W^{+b} W^{+c} D^{+}_f W^{- d}\,, \lb{INV3} \\
&& Y_1 = \varepsilon_{abcd} {\rm Tr} \int dZ\, W^{+ a} W^{+ b} {\cal D}_{fg} W^{-c} {\cal D}^{fg}W^{- d}\,, \lb{INV4} \\
&& Y_2 = \varepsilon_{abcd} {\rm Tr} \int dZ\, W^{+ a} W^{- b} {\cal D}_{fg} W^{-c} {\cal D}^{fg}W^{+ d}\,. \lb{INV4a}
\eea

Using the list of integrals given in Appendix, the explicit expressions for the ${\cal N}=(1,0)$ SYM cores of  the invariants $S^{(1)}_{I, II}$ found earlier
can be expressed through the independent integrals as
\bea
S^{(1)}_{I} \,\rightarrow \, -2\big( E_2 + E_3 + Y_1 \big)\,, \quad S^{(1)}_{II} \,\rightarrow \,-2\big[ 4 E_1 + 3(E_2 + E_3)  + 2( Y_1 +Y_2) \big]. \lb{ComPrev}
\eea
Note that the only combination of $Y_1$ and $Y_2$ which vanishes on shell in the abelian limit is
\be
Y_1 + 2 Y_2\,. \lb{Vanish}
\ee
It appears just in the difference $S^{(1)}_{I} - S^{(1)}_{II}$, which can be transformed on shell to the form $ \sim \int {\rm Tr}\,d\hat{Z}\,[\phi^{+\hat{+}}, \phi^{-\hat{-}}]^2$
that vanishes in the abelian limit (the superfield densities of the structures $E_{1, 2, 3}$ are reduced to total $D^{+}_a$
derivatives in this limit).

Now we can proceed to the remaining invariants $S^{(1)}_{(12)(3)}$  and $S^{(1)}_{(12)(2)}$ defined in \p{AlloneTrace}.

It will be convenient to rewrite, using various on-shell properties of $\phi^{\pm\hat{\pm}}$ and integrating by parts,
the Lagrangian density $L^{(1)}_{(12)(3)}$ as
\bea
L^{(1)}_{(12)(3)} \, \simeq \, - {\rm Tr}\,\big(\phi^{+\hat{+}}\phi^{-\hat{-}} D^{\hat{+} a} \nabla^+_a \phi^{-\hat{-}}\big)\,,
\eea
and also to take into account that
\bea
\nabla^{\pm}_a = {\cal D}^{\pm}_a + \theta^{\hat{-}}_a[\phi^{\pm\hat{+}}, \;]\,. \nonumber
\eea

Rather boring (albeit straightforward) calculations yield
\bea
S^{(1)}_{(12)(3)} \, \rightarrow \, 2\big[8E_1 + 7 E_2 + E_3 + 2 (Y_1 + 2 Y_2)\big]. \lb{12(3)}
\eea
We observe that here just the combination \p{Vanish} appears, so the ${\cal N}=(1,0)$ core of the 3-superfield invariant
is on-shell vanishing in the abelian limit.

One way to calculate the ${\cal N}=(1,0)$ limit of the Lagrangian density of the remaining invariant $S^{(1)}_{(12)(2)}$ is to use the explicit form of the
vector connection \p{VabEx}. Another, simpler way is to note that this Lagrangian can be conveniently rearranged as
\bea
L^{(1)}_{(12)(2)} \, \simeq \, - \frac14 {\rm Tr}\,\big(D^{\hat{+} a}D^{\hat{+} b}\phi^{+\hat{-}} \nabla^-_b \nabla^-_a \phi^{+\hat{-}} \big).
\eea
Once again, rather cumbersome time-consuming calculations finally yield the expression
\bea
S^{(1)}_{(12)(2)} \, \rightarrow \,  4\big(2E_1 + E_2 + E_3) + 2(Y_1 + 2 Y_2)\,, \lb{12(2)}
\eea
from which it follows, in particular, that this invariant also does not contribute to the abelian limit.

Finally, to give a feeling how these invariants look in terms of components, we will present the component form of the abelian ${\cal N}=(1,0)$ superfield
Lagrangian \p{Abel}, with all fields, except for the gauge field $A_{ab}(x)$, being omitted. The corresponding form of $V^{++}$ in WZ gauge and in the analytic basis is just
\bea
V^{++} = i\theta^{+ a}\theta^{+ b} A_{ab}\,. \lb{WZA}
\eea
The relevant non-analytic harmonic connection $V^{--}$ is easily restored from the abelian version of the flatness condition \p{Flat10}:
\bea
V^{--} &=& i\theta^{- a}\theta^{-b} A_{ab} + \Psi^{-3}_d\theta^{+ c} {\cal F}^d_c\,, \quad
{\cal F}^d_c := \frac16 \varepsilon^{abfd}\big(\partial_{ab} A_{fc} - \partial_{fc} A_{ab}\big),
\eea
whence
\bea
W^{+ a} = -\frac16 \varepsilon^{abcd} D^{+}_b D^{+}_c D^{+}_d V^{--} = - \theta^{+ b}{\cal F}^a_b\,.
\eea
That part of $W^{-a} = {D}^{--}W^{+ a}$ which contributes to the Berezin integral in \p{Abel} can also be easily found
\bea
W^{-a} \, \rightarrow \, -i \theta^{-b}\theta^{-c}\theta^{+ d} \partial_{bc}{\cal F}^a_d \,.
\eea
Then it is straightforward to perform the Berezin integration and to find
\bea
S^{\rm (abel)}_{(12)} = -2 \varepsilon^{abcd}\varepsilon_{efgh}
\int d^6 x\, \big(\partial^{lt}{\cal F}^e_a \partial_{lt}{\cal F}^f_b\big)\,\big(\partial^{mn}{\cal F}^g_c \partial_{mn}{\cal F}^h_d \big)\,.\lb{SpinAbel}
\eea

The same expression in the vector notation looks rather bulky.  Passing to the vector notation according to eqs. \p{VecSpin}, \p{Mab}, we obtain\footnote{The gauge potentials $A_M$
are imaginary with our conventions.}
\bea
{\cal F}^a_b = -\frac{1}{12}\,(\sigma^{MN})^a_b {\cal F}_{MN}\,, \quad  {\cal F}_{MN} = \partial_M A_N - \partial_N A_M = \frac32\, {\cal F}^a_b (\sigma_{MN})^b_a\,,
\eea
and rewrite, up to a numerical coefficient, the action \p{SpinAbel} with the help of relations \p{GammaRel} as
\bea
&& S^{\rm (abel)}_{(12)} \sim \int d^6 x \Big[ \big(\partial {\cal F}^{MN} \cdot \partial {\cal F}_{MN}\big)^2
+ 2\big(\partial {\cal F}^{MN}\cdot \partial {\cal F}^{ST}\big)\big(\partial {\cal F}_{MN}\cdot \partial {\cal F}_{ST}\big) \nn
&& -\,4\big(\partial {\cal F}^{MN}\cdot  \partial {\cal F}^{ST}\big)\big(\partial {\cal F}_{MS} \cdot \partial {\cal F}_{NT}\big) -
8\big(\partial {\cal F}^{M}_{\;\;N}\cdot \partial {\cal F}_{MT}\big)\big(\partial {\cal F}^{SN} \cdot \partial {\cal F}_{S}^{\;\;T}\big)\Big], \nonumber
\eea
where $\partial {\cal F}^{MN} \cdot \partial {\cal F}_{MN} := \partial^L {\cal F}^{MN}\partial_L {\cal F}_{MN}$, etc.

It is interesting that there exist no ${\bf d=12}$ invariants which could yield, in the abelian gauge theory limit, the expression like $\sim F^6$, as distinct from the
${\bf d=8}$ invariants which contain $F^4$ term in such a limit \cite{12}. It is likely that the abelian  $F^6$ terms do not admit even ${\cal N}=(1,0)$ completion.


\section{Summary and outlook}
In this paper we continued the applications of the on-shell $6D, {\cal N}=(1,1)$ HSS for constructing higher-derivative invariants of the maximally
supersymmetric $6D$ SYM theory. We have explicitly constructed new, dimension ${\bf d=12}$, single-trace invariants \p{AlloneTrace} and the double-trace invariants \p{All2Trace},
studied their ${\cal N}=(1,0)$ SYM cores and the component gauge-field structure of their abelian limit. All these invariants are given by integrals over the total bi-harmonic
$6D$ superspace and we did not find any ``no-go''-type conjectures which would forbid
the appearance of the four-loop logarithmic divergences of such a form within the off-shell $6D, {\cal N}=(1,0)$ HSS supergraph calculations
in the quantum ${\cal N}=(1,1)$ SYM theory. The leading in momenta contributions should come from \p{All2Trace} and the invariants in the first line of \p{AlloneTrace}, while the rest of
the single-trace invariants, including additional derivatives, should be responsible for some sub-leading terms (divergent and/or finite) in
the relevant order of the total effective action.

It would be interesting to look for the invariants of the dimension ${\bf d=14}$ and ${\bf d=16}$ corresponding to the five- and six-loop contributions. The leading terms
should be a direct generalization of those in \p{AlloneTrace} and \p{All2Trace}. E.g.,  one of the possible candidates for such invariants of the dimension ${\bf d=14}$ reads
\bea
 \sim {\rm Tr} \int d\hat{Z}\, (\phi^{+\hat{+}})^3 (\phi^{-\hat{-}})^3\,, \nonumber
\eea
and it seems that finding the full set of independent on-shell invariants of this sort could be a comparatively easy straightforward task. On the other hand, according to
the arguments of ref. \cite{Smi}, already at the dimension ${\bf d=16}$ there should be taken into account deformations of the conventional mass shell associated with the
superfield equations of motion \p{EqMV+q} and  \p{EqMq}. So the formalism of the on-shell $6D, {\cal N}=(1,1)$ HSS should undergo a proper modification. It   would be extremely interesting
to explicitly learn how this would happen, how the initial constraints of $6D, {\cal N}=(1,1)$ SYM theory are to be corrected and which geometric principle could stand behind
such a modification.

\section*{Acknowledgements}
E.I. thanks G. Bossard, I. Buchbinder and A. Smilga for illuminating discussions of various issues related to the subject of
this paper.  The work of E.I. was partly supported by Russian Scientific Foundation, project No 21-12-00129.

\bigskip

\renewcommand\theequation{A.\arabic{equation}} \setcounter{equation}0
\subsection*{A\quad Some integrals }
Here we express, through the basic structures defined in \p{INV1} - \p{INV3}, various ${\cal N}=(1,0)$ integrals encountered when calculating the ${\cal N}=(1,0)$ SYM
limit of the ${\cal N}=(1,1)$ invariants in Sect. 7. While transforming the relevant integrands, we integrate by parts and use the on-shell
conditions $D^+_a W^{ + d} = {\cal D}^-_a W^{- d} = 0\,, D^+_a W^{ - d} = - {\cal D}^-_a W^{+ d}\,, D^+_a W^{ - a} = 0\,, \nabla^{\pm\pm} W^{\pm a}= 0\,,
\nabla^{\pm\pm} W^{\mp a} = W^{\pm a}$. In this way we derive the following list of the on-shell ${\cal N}=(1,0)$ SYM integrals
\bea
&&\varepsilon_{abcd}\,{\rm Tr} \int dZ\, W^{+ a}W^{- f}W^{+ b} W^{- c} D^+_f W^{- d} = -(E_1 + E_2 + E_3)\,, \nn
&& \varepsilon_{abcd}\,{\rm Tr} \int dZ\,  W^{+ a}W^{- f}W^{- b} W^{+ c} D^+_f W^{- d} = E_1 + E_2\,, \nn
&&\varepsilon_{abcd}\,{\rm Tr} \int dZ \, W^{+ a}W^{- b}W^{+ c}W^{- f} D^+_f W^{- d} = - (E_1 + E_2)\,, \nn
&& \varepsilon_{abcd}\,{\rm Tr} \int dZ \, W^{- a}W^{+ b}W^{- f} W^{+ c} D^+_f W^{- d}  = -(E_1 + E_2 + E_3)\,, \nn
&&\varepsilon_{abcd}\,{\rm Tr} \int dZ \, W^{- f}  W^{+a}W^{+ b} W^{- c} D^+_f W^{- d}  = E_3\,, \nn
&& \varepsilon_{abcd}\,{\rm Tr} \int dZ \, W^{- f}  W^{-a}W^{+ b} W^{+ c} D^+_f W^{- d}  = E_1\,. \lb{IntN10}
\eea



\begin{thebibliography}{99}
\bibitem{Tseytlin}
  A.A.~Tseytlin, {\it On non-abelian generalization of Born-Infeld action in string theory},
  Nucl.\ Phys.\ B {\bf 501} (1997) 41-52,
  {\tt arXiv:hep-th/9701125}.
\bibitem{3} C. Cheung and D. O'Connell, {\it Amplitudes and Spinor-Helicity in Six
Dimensions}, JHEP {\bf 0907} (2009) 075,  {\tt arXiv:0902.0981 [hep-th]}.
\bibitem{4} J. Plefka, T. Schuster, and V. Verschinin, {\it From Six to Four and More:
Massless and Massive Maximal Super Yang-Mills Amplitudes in 6d and 4d and their
Hidden Symmetries},  JHEP {\bf 1501} (2015) 098, {\tt arXiv:1405.7248 [hep-th]}.
\bibitem{5}  T. Dennen, Yu-tin Huang, and W. Siegel, {\it Supertwistor space for 6D
maximal super Yang-Mills},  JHEP {\bf 1004} (2010) 127,  {\tt arXiv:0910.2688 [hep-th]}.

\bibitem{6} Yu-tin Huang and A.E. Lipstein, {\it Amplitudes of 3D and 6D Maximal
Superconformal Theories in Supertwistor Space},  JHEP {\bf 1010} (2010) 007,  {\tt arXiv:1004.4735 [hep-th]}.
\bibitem{7} Z. Bern, J.J. Carrasco, T. Dennen, Yu-tin Huang, and H. Ita,
{\it Generalized Unitarity and Six-Dimensional Helicity}, Phys. Rev. D {\bf 83} (2011) 085022, {\tt arXiv:1010.0494 [hep-th]}.

\bibitem{9} Z. Bern, J.J.M. Carrasco, and H. Johansson, {\it New Relations for Gauge-Theory
Amplitudes},  Phys. Rev. D {\bf 78} (2008) 085011,  {\tt arXiv:0805.3993 [hep-ph]}.
\bibitem{13} Z. Bern, L.J. Dixon, and V.A. Smirnov, {\it Iteration of planar amplitudes
in maximally supersymmetric Yang-Mills theory at three loops and beyond}, Phys. Rev. D {\bf 72} (2005) 085001, {\tt arXiv:hep-th/0505205}.

\bibitem{14} Z. Bern, J.J.M. Carrasco, L.J. Dixon, H. Johansson, and R. Roiban,  {\it The
Complete Four-Loop Four-Point Amplitude in N=4 Super-Yang-Mills Theory}, Phys. Rev. D {\bf 82} (2010) 125040,  {\tt arXiv:1008.3327 [hep-th]}.
\bibitem{Bern:2012uf}
  Z.~Bern, J.~J.~M.~Carrasco, L.~J.~Dixon, H.~Johansson and R.~Roiban,
{\it Simplifying Multiloop Integrands and Ultraviolet Divergences of Gauge Theory
  and Gravity Amplitudes},
  Phys.\ Rev.\ D {\bf 85} (2012) 105014, {\tt arXiv:1201.5366 [hep-th]}.

\bibitem{BGRV}
N.~Berkovits, M.~B.~Green, J.~G.~Russo and P.~Vanhove,
{\it Non-renormalization conditions for four-gluon scattering in supersymmetric string and field theory},
JHEP {\bf 0911} (2009) 063, {\tt arXiv:0908.1923 [hep-th]}.

\bibitem{Bjornsson:2010wm}
J.~Bjornsson and M.~B.~Green, {\it 5 loops in 24/5 dimensions},
JHEP {\bf 1008} (2010) 132, {\tt arXiv:1004.2692 [hep-th]}.

\bibitem{Bjornsson:2010wu}
J.~Bjornsson, {\it Multi-loop amplitudes in maximally supersymmetric pure spinor field theory},
JHEP {\bf 1101} (2011) 002, {\tt arXiv:1009.5906 [hep-th]}.


\bibitem{BKKTV} L.V. Bork, D.I. Kazakov, M.V. Kompaniets, D.M. Tolkachev, and D.E. Vlasenko, {\it Divergences in maximal supersymmetric
Yang-Mills theories in diverse dimensions}, JHEP {\bf 1511} (2015) 059,  {\tt arXiv:1508.05570 [hep-th]}.
\bibitem{25} G. Bossard, P.S. Howe, and K.S. Stelle, {\it The Ultra-violet question in maximally
supersymmetric field theories},  Gen. Rel. Grav. {\bf 41} (2009) 919-981, {\tt arXiv:0901.4661 [hep-th]}.
\bibitem{Bossard:2009mn} G.~Bossard, P.S.~Howe and K.S.~Stelle, {\it A Note on the UV behaviour of maximally supersymmetric Yang-Mills theories},
Phys.\ Lett.\ B {\bf 682} (2009) 137, {\tt arXiv:0908.3883 [hep-th]}.

\bibitem{16} P.S. Howe, G. Sierra, and P.K. Townsend, {\it Supersymmetry in Six-Dimensions},
Nucl. Phys. B {\bf 221} (1983) 331-348.
\bibitem{Koller} J.~Koller, {\it A six-Dimensional superspace approach to extended superfioelds}, Nucl. Phys. B {\bf 222} (1983) 319.
\bibitem{20} A. Galperin, E. Ivanov, S. Kalitzyn, V. Ogievetsky, and E. Sokatchev, {\it Unconstrained
N=2 Matter, Yang-Mills and Supergravity Theories in Harmonic Superspace}, Class.
Quant. Grav. {\bf 1} (1984) 469-498 [Erratum: Class. Quant. Grav. {\bf 2} (1985) 127].
\bibitem{18} A.S. Galperin, E.A. Ivanov, V.I. Ogievetsky, and E.S. Sokatchev,
{\it Harmonic superspace}, Cambridge Monographs on Mathematical Physics,
Cambridge University Press, 2007, 309 p.
\bibitem{19} P.S. Howe, K.S. Stelle, and P.C. West,  {\it N=1, d = 6 harmonic
superspce}, Class. Quant. Grav. {\bf 2} (1985) 815.
\bibitem{22} B.M. Zupnik, {\it Six-dimensional Superguge Theories in the Harmonioc Superspace}, Sov. J. Nucl. Phys. {\bf 44} (1986) 512 [Yad. Fiz. {\bf 44} (1986) 794-802].
\bibitem{11} P.S. Howe and K. S. Stelle, {\it  Ultraviolet Divergences in Higher Dimensional
Supersymmetric Yang-Mills Theories},  Phys. Lett. B {\bf 137} (1984) 175-180.
\bibitem{11a} P.S. Howe and K. S. Stelle, {\it Supersymmetry counterterms revisited}, Phys. Lett. B {\bf 554} (2003) 190, {\tt arXiv:hep-th/0211279}.
%
\bibitem{Int}
G. Bossard, P.S. Howe, U. Lindstrom, K.S. Stelle, L. Wulff, {\it Integral invariants in maximally supersymmetric Yang-Mills theories},
JHEP {\bf 1105} (2011) 021, {\tt arXiv:1012.3142 [hep-th]}.
\bibitem{12} G. Bossard, E. Ivanov, and A. Smilga, {\it Ultraviolet behavior of 6D
supersymmetric Yang-Mills theories and harmonic superspace}, JHEP {\bf 1512} (2015) 085, {\tt arXiv:1509.08027 [hep-th]}.

\bibitem{BuIvIv} I.L. Buchbinder, E.A. Ivanov, and S.A. Ivanovskiy, {\it New bi-harmonic superspace formulation of $4D, {\cal N}=4$ SYM theory}, JHEP {\bf 2104} (2021) 010,
{\tt arXiv:2012.09669 [hep-th]}.

\bibitem{Smi} A. Smilga, {\it Ultarviolet divergences in non-renormalizable supersymmetric theories}, Phys. Part. Nucl. Lett. {\bf 14} (2017) no.2, 245-260,
{\tt arXiv:1603.06811 [hep-th]}.
\bibitem{Q6D1} I.L. Buchbinder, E.A. Ivanov, B.S. Merzlikin, and K.V. Stepanyantz, {\it One-loop divergences in the $6D, \,{\cal N}=(1,0)$
abelian gauge theory}, Phys. Lett. B {\bf 763} (2016) 375, {\tt arXiv:1609.00975 [hep-th]}.
\bibitem{Q6D2} I.L. Buchbinder, E.A. Ivanov, B.S. Merzlikin, and K.V. Stepanyantz, {\it One-loop divergences in $6D, \,{\cal N}=(1,0)$
SYM theory}, JHEP {\bf 1701} (2017) 128, {\tt arXiv:1612.03190 [hep-th]}.
\bibitem{Q6D3} I.L. Buchbinder, E.A. Ivanov, B.S. Merzlikin, and K.V. Stepanyantz, {\it Supergraph analysis of the one-loop divergences
in $6D, \,{\cal N}=(1,0)$ and ${\cal N}=(1,1)$ SYM theories}, Nucl. Phys. B {\bf 921} (2017) 127, {\tt arXiv:1704.02530 [hep-th]}.

\bibitem{17}  E.A. Ivanov, A.V. Smilga, and B.M. Zupnik, {\it Renormalizable supersymmetric
gauge theory in six dimensions},  Nucl. Phys. B {\bf 726} (2005) 131-148, {\tt arXiv:hep-th/0505082}.

\bibitem{24} I.L. Buchbinder and N.G. Pletnev, {\it Construction of $6D$ supersymmetric field models
in ${\cal N}=(1,0)$ harmonic superspace}, Nucl. Phys. B {\bf 892} (2015) 21-48,  {\tt arXiv:1411.1848 [hep-th]}.

\bibitem{Kuz1} S.M.~Kuzenko, J.~Novak, and I.B.~Samsonov, {\it The anomalous current multiplet in 6D minimal supersymmetry}, JHEP {\bf 1602} (2016) 132 {\tt arXiv:1511.06582 [hep-th]}.
\bibitem{Kuz2} S.M.~Kuzenko, J.~Novak, and I.B.~Samsonov, {\it Chiral anomalies in six dimensions from harmonic superspce}, JHEP {\bf 1711} (2017) 145, {\tt arXiv:1708.08238 [hep-th]}.
\bibitem{2La} I.L.~Buchbinder, E.A.~Ivanov, B.S.~Merzlikin, and K.V.~Stepanyantz,
  {\it On the two-loop divergences of the 2-point hypermultiplet supergraphs for $6D$, ${\cal N} = (1,1)$ SYM theory},
  Phys.\ Lett.\ B {\bf 778} (2018) 252, {\tt arXiv:1711.11514 [hep-th]}.
\bibitem{2Lb} I.L.~Buchbinder, E.A.~Ivanov, B.S.~Merzlikin, and K.V.~Stepanyantz,
  {\it On the two-loop divergences in $6D$, ${\cal N} = (1,1)$ SYM theory}, Phys.\ Lett.\ B {\bf 820} (2021) 136516, {\tt arXiv:2104.14284 [hep-th]}.

\bibitem{CS1} A.A.~Tseytlin,  K.~Zarembo,
 {\it  Magnetic interactions of D-branes and Wess-Zumino terms in superYang-Mills
effective actions}, Phys. Lett.  B {\bf 474} (2000) 95-102, {\tt  arXiv:hep-th/9911246}.

\bibitem{CS2}
 K.A.~Intriligator,  {\it Anomaly matching and a Hopf-Wess-Zumino term in
$6d, {\cal N}=(2,0)$ field theories},  Nucl. Phys.  B {\bf 581} (2000) 257-273, {\tt  arXiv:hep-th/0001205}.

\bibitem{CS3}P.C.~Argyres, A. M.~Awad,  G. A.~Braun,   F.P.~Esposito,
{\it Higher-Derivative Terms in ${\cal N}=2$ Supersymmetric Effective
Actions}, JHEP {\bf 0307} (2003) 060, {\tt  arXiv:hep-th/0306118}.


\end{thebibliography}
\end{document}